\documentclass[
superscriptaddress,
amsmath,amssymb,
aps,
prb,
twocolumn, floatfix, reprint,
longbibliography
]{revtex4-2}

\usepackage{graphicx}
\usepackage{dcolumn}
\usepackage{bm}

\usepackage{tensor}
\usepackage{accents}
\usepackage[shortlabels]{enumitem}
\usepackage[caption=false]{subfig} 
\usepackage{xcolor}

\usepackage[colorlinks,urlcolor=blue,linkcolor=blue,citecolor=blue,breaklinks=True]{hyperref} 
\usepackage[capitalize,noabbrev]{cleveref} 

\newcommand{\ket}[1]{\left\lvert #1 \right\rangle} 
\newcommand{\outerprod}[2]{{\left\vert #1 \vphantom{#2} \right\rangle} {\left\langle #2 \vphantom{#1} \right\vert}} 

\begin{document}

\preprint{APS/123-QED}

\title{Long-lived oscillations of metastable states in neutral atom systems}

\author{Siva~Darbha}
 \email{sdarbha@lbl.gov}
 \email{siva.darbha@berkeley.edu}
 \affiliation{Applied Mathematics and Computational Research Division, Lawrence Berkeley National Laboratory, Berkeley, CA 94720, USA}
 \affiliation{National Energy Research Scientific Computing Center, Lawrence Berkeley National Laboratory, Berkeley, CA 94720, USA}

\author{Milan~Kornja\v{c}a}
 \affiliation{QuEra Computing Inc., 1284 Soldiers Field Road, Boston, MA 02135, USA}

\author{Fangli~Liu}
 \affiliation{QuEra Computing Inc., 1284 Soldiers Field Road, Boston, MA 02135, USA}

\author{Jan~Balewski}
 \affiliation{National Energy Research Scientific Computing Center, Lawrence Berkeley National Laboratory, Berkeley, CA 94720, USA}

\author{Mark~R.~Hirsbrunner}
 \affiliation{National Energy Research Scientific Computing Center, Lawrence Berkeley National Laboratory, Berkeley, CA 94720, USA}
 \affiliation{Department of Physics and Institute for Condensed Matter Theory, University of Illinois at Urbana-Champaign, Urbana, IL 61801, USA}

\author{Pedro~L.~S.~Lopes}
 \affiliation{QuEra Computing Inc., 1284 Soldiers Field Road, Boston, MA 02135, USA}

\author{Sheng-Tao~Wang}
 \affiliation{QuEra Computing Inc., 1284 Soldiers Field Road, Boston, MA 02135, USA}

\author{Roel~Van~Beeumen}
 \affiliation{Applied Mathematics and Computational Research Division, Lawrence Berkeley National Laboratory, Berkeley, CA 94720, USA}

\author{Katherine~Klymko}
 \affiliation{National Energy Research Scientific Computing Center, Lawrence Berkeley National Laboratory, Berkeley, CA 94720, USA}

\author{Daan~Camps}
 \affiliation{National Energy Research Scientific Computing Center, Lawrence Berkeley National Laboratory, Berkeley, CA 94720, USA}


\begin{abstract}
Metastable states arise in a range of quantum systems and can be observed in various dynamical scenarios, including decay, bubble nucleation, and long-lived oscillations. The phenomenology of metastable states has been examined in quantum many-body systems, notably in 1D ferromagnetic Ising spin systems and superfluids. In this paper, we study long-lived oscillations of metastable and ground states in 1D antiferromagnetic neutral atom chains with long-range Rydberg interactions. We use a staggered local detuning field to achieve confinement. Using theoretical and numerical models, we identify novel spectral signatures of quasiparticle oscillations distinct to antiferromagnetic neutral atom systems and interpret them using a classical energy model of short-range meson repulsion. Finally, we evaluate the experimental accessibility of our proposed setup on current neutral-atom platforms and discuss experimental feasibility and constraints.
\end{abstract}

\maketitle

\section{Introduction}
\label{sec:intro}

Metastable states are excited states that can arise in quantum field theory and quantum many-body physics, separated in energy from the global ground state. These states were explored in statistical physics~\cite{Langer:1967,Lifshitz:1972}, and then prominently in quantum field theory in the form of false vacua~\cite{Kobzarev:1974cp,Stone:1976,Stone:1977,Coleman:1977a,Coleman:1977b,Callan:1977,Voloshin:1985id}. Metastable and ground states can be identified and distinguished by the signatures they produce during dynamical processes, including decay, bubble nucleation, and long-lived oscillations. The detection of these states can serve as a probe of the structure of the underlying system and, by extension, its history. 

In quantum field theory, pioneering studies characterized the decay of false vacua for applications in early-Universe cosmology and fundamental interactions~\cite{Kobzarev:1974cp,Stone:1976,Stone:1977,Coleman:1977a,Coleman:1977b,Callan:1977,Coleman:1980,Voloshin:1985id}. The decay process occurs by tunneling via bubble nucleation that precipitates a first order phase transition~\cite{Coleman:1977a,Coleman:1977b}, and can be summarized as follows. Quantum fluctuations induce the nucleation of true vacuum domains in the false vacuum background. The fate of a domain is determined by its size, according to the energy trade-off between the bulk gain and surface cost; there is a critical size above which it is energetically favorable for the domain to expand. These supercritical domains expand throughout the system, whereas subcritical ones cannot. 
However, for certain system parameters, the tunneling probability can be extremely small, which would lead to a decay process that only unravels over an extremely large timescale. In cosmology, the timescale could be larger than the age of the Universe (e.g.~\cite{Mazumdar:2019,Hindmarsh:2021,Athron:2024}).

Phase transitions with metastable states are also prominent processes in quantum many-body systems. In ultracold atomic systems, a number of experiments have studied such phase transitions~\cite{Hruby:2018,Song:2022,Zenesini:2024}, including recent work that detected metastable decay using ferromagnetic superfluids formed from atomic Bose-Einstein condensates~\cite{Zenesini:2024}, and many others have been proposed (e.g.~\cite{Opanchuk:2013,Fialko:2015,Fialko:2017,Jenkins:2023,Jenkins:2024} and references therein). In 1D Ising spin systems, an array of analytic and numerical efforts have probed phase transitions involving metastability, aided by an improved understanding of confinement physics in these systems~\cite{Lake:2010,Kormos:2017,Liu:2019,Tan:2021}. The rate of decay has been derived and quantified in detail~\cite{Rutkevich:1999,Lagnese:2021}. Numerical studies have shown that a linear ramp across a first order phase transition can produce nucleation events in the background state at quantized values of the ramp field~\cite{Sinha:2021}. Simulations of domain wall collisions have shown that confined domains can be produced in these scattering events~\cite{Milsted:2022}.

Instead of phase transitions, metastable and ground states can also be detected through long-lived oscillations arising from superpositions with excited states containing subcritical domains~\cite{Pomponio:2022,Lagnese:2023}. In contrast to decay, long-lived oscillations are readily observable and do not depend on a vanishingly small occurrence probability. In 1D Ising spin chains, recent numerical work has examined the Fourier spectra of oscillations after a quantum quench, finding spectral trends that can be used to identify and discriminate between metastable and ground states~\cite{Lagnese:2023}. This followed earlier work that observed Bloch oscillations of domain walls after a quench, producing a regular peak structure in the Fourier spectra~\cite{Pomponio:2022}. In neutral atom chains, oscillations have been briefly examined in a supplementary study, under some simplifying conditions~\cite{Lagnese:2023}. Long-lived oscillations can also arise in quantum many-body systems under conditions unrelated to metastability, and can be similarly used to interrogate the system properties (e.g.~\cite{Robertson:2024} and references therein).

False and true vacuum phenomenology in quantum field theory is conceptually similar to metastable and ground state phenomenology in truncated quantum many-body systems, but differs fundamentally since the vacuum state of a quantum field includes all field modes whereas the states treated in a finite-dimensional Hilbert space are comparatively restricted. Metastable states have intrinsically interesting dynamics in condensed matter systems, and though observations on these systems cannot be directly applied to quantum fields, they can interrogate the emergence of new phenomena in truncated systems and potentially inspire further theoretical studies.

Neutral atoms have recently become a prominent platform for the study of quantum many-body physics in the antiferromagnetic phase~\cite{Browaeys:2020}, and have been implemented as qubits in digital and analog quantum hardware~\cite{Isenhower:2010,Saffman:2010,Henriet:2020,aquila2023quera}. Notably, pioneering efforts have developed techniques to engineer strong interactions using Rydberg excitations~\cite{Jaksch:2000,Lukin:2001,Gaetan:2009,Urban:2009} and to control and arrange individual atoms using optical tweezers~\cite{Nogrette:2014,Labuhn:2014,Kaufman:2014,Barredo:2016,Endres:2016,Labuhn:2016,Scholl:2021,Ebadi:2021}. Neutral atom arrays in one and two dimensions have motivated many novel investigations, such as: the discovery of quantum many-body scars~\cite{Bernien:2017,Turner:2018,Bluvstein:2021}; the corroboration of the quantum Kibble-Zurek mechanism~\cite{Keesling:2019,Ebadi:2021}; the computation of ground state phase diagrams~\cite{Samajdar:2020,Samajdar:2021,Ohler:2022,Chen:2023}; the exploration of peculiar phases~\cite{Lienhard:2020,Zhang:2024}; the study of topological phases~\cite{deLeseleuc:2019,Verresen:2021,Semeghini:2021,Ohler:2023,Kornjaca:2023}; the real-time simulation of string dynamics and scattering in lattice gauge theories~\cite{Surace:2020,Surace:2021}; and the solutions of Maximum Independent Set problems on unit disk graphs~\cite{Ebadi:2022,Nguyen:2023,Finzgar:2023}.

We study the long-lived oscillations of metastable and ground states in 1D neutral atom chains mediated by Rydberg interactions. Our main tools are numerical simulations and analytic modeling. We perform a broad investigation of the parameter space and include long-range Rydberg tails, distinct from the supplementary study of Ref.~\cite{Lagnese:2023}. Our setups are accessible to near-term experiments. We determine the spectral features of quasiparticle oscillations involving metastable and ground states. We identify features distinct to these systems and explain them through the competition between confinement and short-range repulsive effects. Our results show that the detailed microphysics of the system influence the spectral features, and thus the ability to distinguish between metastable and ground states.

In a concurrent work, we study the decay and nucleation dynamics of metastable excited states in neutral atom systems~\cite{Darbha:2024a}. The parameter regime for that process is complementary to the one examined in this paper. 

The remainder of our paper is organized as follows. We describe our model in \Cref{sec:model}. We summarize some theoretical background for modeling the dynamics of single bubbles in \Cref{sec:quasiparticle}. Our numerical setup and results are presented in \Cref{sec:numerical}. We interpret our results in terms of a short-range meson repulsion model in \Cref{sec:repulsion}. In \Cref{sec:experiments}, we discuss the experimental accessibility of our setups and results. We conclude and present our outlook in \Cref{sec:conclusion}.
\section{Model}
\label{sec:model}

\begin{figure}
\centering
\includegraphics[width=0.48\textwidth]{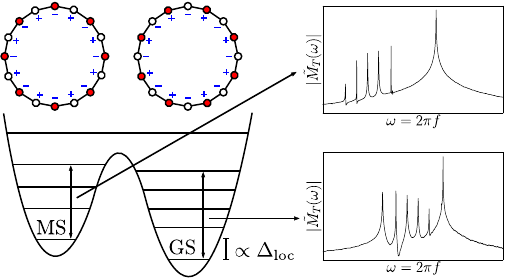}
\caption{A schematic overview of our system and results. The atoms are arranged in a circle with 16 sites and constant atom separation. The blue $+/-$ signs show the spatial pattern of the staggered local detuning field $\Delta_{\mathrm{loc},j}$ that we apply from \Cref{eq:local_detuning}, here shown for $\Delta_\mathrm{loc} > 0$. The field produces an asymmetric potential with metastable state (MS) and ground state (GS) separated by an energy density $\mathcal{E} \propto \Delta_\mathrm{loc}$. The states are dominated by the two $Z_2$ product states, shown in the chains above, where a red atom indicates a Rydberg state $\ket{r}$ and a white atom indicates a ground state $\ket{g}$. The metastable state undergoes coherent oscillations with low-lying excited states, only schematically shown here inside the left well. The magnetization is a suitable observable, defined in Equations~\eqref{eq:magnetization} and \eqref{eq:M_T}. Its Fourier spectrum is shown in the top right panel, exhibiting a distinctive peak structure. Similarly, the ground state undergoes coherent oscillations, and the associated spectrum is shown in the bottom right panel. The peak structure depends on the microscopic details of the confining potential. It can be used to distinguish between the two states, and understood from our analysis in \Cref{sec:repulsion}.}
\label{fig:overview}
\end{figure}

We study long-lived oscillations of an initial state near the metastable and ground states. \Cref{fig:overview} schematically shows the dynamics and signatures in these two cases, which will be elucidated in Sections \ref{sec:quasiparticle} and \ref{sec:numerical}. In our numerical studies, we set the initial state near one of the states and track its evolution using various observables. In this section, we describe our system and observables. 

We examine a 1D neutral atom chain with an even number of sites $n_s$, setting $n_s = 16$, constant atom separation $a$, and a circle geometry yielding periodic boundary conditions (\Cref{fig:overview}). Each atom $j$ is a two-level system with a ground state $\ket{g_j} \equiv \ket{0_j}$ and a Rydberg state $\ket{r_j} \equiv \ket{1_j}$. A state of the full chain can be expanded in terms of the site (or computational) basis $\ket{x}$ where $x \in \{0,1\}^{\otimes n_s}$, where the leftmost bit labels atom 1 and the rightmost bit labels atom $n_s$. 

The Hamiltonian is
\begin{equation}
H = \frac{\Omega}{2} \sum_j \hat{\sigma}_{x,j} - \sum_j \Delta_j \hat{n}_j + \sum_{j<k} V_{jk} \hat{n}_j \hat{n}_k \, ,
\label{eq:hamiltonian}
\end{equation}
where $\hat{\sigma}_{x,j} = \outerprod{g_j}{r_j} + \outerprod{r_j}{g_j}$ is the Pauli-$\hat{X}$ operator, $\hat{n}_j = \outerprod{r_j}{r_j}$ is the number operator, $\Omega$ is the Rabi frequency that induces atomic transitions between the ground and Rydberg state, $\Delta_j$ is the detuning field, and $V_{jk} = C_6 / r_{jk}^6$ is the Rydberg-Rydberg interaction potential mediated by van der Waals interactions, with $C_6$ a constant determined by the atom species and $r_{jk}$ the atom separation. We define $V_1 = C_6 / r_{i,i+1}^6$ and $V_2 = C_6 / r_{i,i+2}^6$ as the nearest-neighbor and next-nearest-neighbor interactions, respectively. The detuning field can be written as
\begin{equation}
\Delta_j = \Delta_\mathrm{glob} + \Delta_{\mathrm{loc},j} \, ,
\end{equation}
i.e.~with global and local components. Importantly, the atoms experience a Rydberg blockade for separations $r_{jk} \lesssim R_b$, where $R_b = (C_6 / \Omega)^{1/6}$ is the blockade radius, within which dual excitations to the Rydberg states are energetically unfavorable due to the severe scaling $V_{jk} \propto 1/r_{jk}^6$~\cite{Jaksch:2000,Lukin:2001}.

We set the Rabi frequency to $\Omega/2\pi = 1.0$ MHz, which simply sets an overall energy scale. The dimensionless ratio $R_b / a$ is tuned by the atom separation $a$, which we choose to be in the range $R_b / a \in (1,2)$ to achieve a nearest-neighbor blockade. We use a constant global detuning field $\Delta_\mathrm{glob}$ to access the $Z_2$ region in the ground state phase diagram~\cite{Bernien:2017,Keesling:2019,Bluvstein:2021}. For $\Delta_\mathrm{glob} \gg \Omega$, the ground state is two-fold degenerate; the degenerate eigenvectors are the two antiferromagnetic $Z_2$ states, adiabatically connected to the product states $\ket{10 \hdots 10}$ and $\ket{01 \hdots 01}$. If we apply a staggered local detuning field
\begin{equation}
\Delta_{\mathrm{loc},j} = (-1)^j \Delta_{\mathrm{loc}} \, ,
\label{eq:local_detuning}
\end{equation}
then we split the ground state into a metastable and ground state separated by energy density $\mathcal{E} \propto \Delta_{\mathrm{loc}}$ (\Cref{fig:overview}). 
We can specify the detuning using two dimensionless free parameters
\begin{align}
\alpha &= \Delta_\mathrm{glob} / \Omega , \label{eq:alpha} \\
\beta &= \Delta_\mathrm{loc}/\Delta_\mathrm{glob} , \label{eq:beta}
\end{align}
where we take $\alpha > 0$. The $Z_2$ product state $\ket{10 \hdots 10}$ will be near the metastable state for small $\beta > 0$ and near the ground state for small $\beta < 0$. \Cref{fig:overview} schematically shows the dynamics and signatures for an initial state in each case, though we note that the approach of the figure is the reverse of the approach we just described, in that the figure shows only $\beta > 0$ and instead selects different initial states to achieve proximity to the metastable and ground states.

The staggered field introduces a confining potential between domain walls that scales linearly with the domain size. The antiferromagnetic domain walls correspond to adjacent $00$ or $11$ configurations, though $11$ domain walls are energetically suppressed when the nearest-neighbor Rydberg blockade is large. The domain, or bubble, size is given by the distance between its domain walls. Domains of even size are suppressed due to the nearest-neighbor Rydberg blockade. The dynamics of domain walls can be effectively described in terms of the dynamics of quasiparticles, using concepts borrowed from high-energy physics. Domain walls can be understood as quarks and antiquarks, depending on the transition direction. In the staggered field, metastable state domains experience a confining interaction potential and thus correspond to mesons, i.e. bound two-quark quasiparticles; ground state domains experience an anti-confining interaction potential. A state with domains contains at least one meson state, possibly more.

The staggered local detuning field explicitly breaks the 1-site translation symmetry of the Hamiltonian $H$, but preserves the 2-site translation symmetry. In other words, the unitary 2-site translation operator $T_2$ preserves the Hamiltonian $H$ under confinement, i.e. $[H,T_2] = 0$ so that $H$ and $T_2$ admit simultaneous eigenstates. The operator $T_2$ is fully characterized by its action
\begin{equation}
T_2 \ket{x} = \ket{x \texttt{>>} 2} \, ,
\end{equation}
on computational basis states $\ket{x}$, where $\texttt{>>}$ is the right bit shift operation. We can write the eigenvalues of $T_2$ as $e^{ik}$, where $k$ is the momentum, as a consequence of Bloch's theorem. We can use the joint eigenvalues of $H$ and $T_2$ to classify the simultaneous eigenstates in the excitation band manifolds.

Globally, an evolved state can be characterized by its magnetic structure, which we describe using order parameters (OPs). The global magnetization $M = \langle \hat{M} \rangle$ describes the ferromagnetic order by the operator
\begin{equation}
\hat{M} = \frac{2}{n_s} \sum_j \left( \frac{1}{2} \hat{I}_j - \hat{n}_j \right) = \frac{1}{n_s} \sum_j \hat{\sigma}_{z,j} \, ,
\label{eq:magnetization}
\end{equation}
where $\hat{\sigma}_{z,j} = \outerprod{g_j}{g_j} - \outerprod{r_j}{r_j}$ is the Pauli-$\hat{Z}$ operator. It equals $M = 0$ for the two $Z_2$ product states, $M = 1$ for the state $\ket{0 \hdots 0}$, and $M = -1$ for the state $\ket{1 \hdots 1}$. For convenience, we further define the transformed magnetization
\begin{equation}
M_T = M - \bar{M} \, ,
\label{eq:M_T}
\end{equation}
which simply subtracts the mean $\bar{M}$. Locally, an evolved state has domain walls separating ground and metastable states. The density of domain walls $\rho = \langle \hat{R} \rangle$ is determined from the operator
\begin{equation}
\hat{R} = \frac{1}{n_s} \sum_j \left[ \hat{n}_j \hat{n}_{j+1} + (\hat{I}_j - \hat{n}_j) (\hat{I}_{j+1} - \hat{n}_{j+1}) \right] \, .
\label{eq:dens_domain_walls}
\end{equation}
The product $\rho n_s$ then gives the real-valued number of domain walls. It equals $\rho n_s = 0$ for the two $Z_2$ product states and $\rho n_s = n_s / 2$ for the two states $\ket{0 \hdots 0}$ and $\ket{1 \hdots 1}$. In the quasiparticle picture, this observable captures the density of mesons.
\section{Quasiparticle Oscillations}
\label{sec:quasiparticle}

In condensed matter systems with confinement, excited states contain mesons, i.e. bound two-quark quasiparticles. A quantum quench can generate a superposition between the ground state and a band of mesons, and the subsequent oscillations encode the properties of the mesons and the confining potential~\cite{Kormos:2017,Liu:2019,Tan:2021}. For instance, the Fourier spectrum of the oscillations has peaks located at the meson masses (for ground state energy set to zero) and their differences~\cite{Kormos:2017,Liu:2019}. For low-energy excitations, the subspace of low-energy single mesons, i.e. two domain walls, can be sufficient to describe the oscillation dynamics and the spectral peaks~\cite{Liu:2019}.

Metastable and ground states can be detected and characterized using an analogous procedure based on these insights. In particular, a quench tuned to either state will produce a superposition between that state and proximate low-energy single meson states. The post-quench oscillations will yield a Fourier spectrum that encodes the metastable or ground state properties through the interference with the meson states. The dynamics are more easily isolated when supercritical bubbles are energetically suppressed and the mesons are subcritical.

In our numerical studies in \Cref{sec:numerical}, we initialize the state near the metastable or ground state, evolve it under the Hamiltonian, and examine the Fourier spectra of the magnetization oscillations. In this section, we present a perturbative model for examining quasiparticle excitations that provides an interpretive framework for our numerical results. We first review earlier work in spin systems. We then present our efforts with neutral atoms. 

\subsection{Spin Chains}
\label{subsec:quasiparticle:spins}

Metastable oscillations have previously been examined using quench dynamics in the ferromagnetic 1D Ising spin model. Notably, in an extensive study, Lagnese et al.~\cite{Lagnese:2023} performed quenches to initial states near either the metastable or ground state, and examined the Fourier spectrum of the magnetization for the resulting long-lived oscillations. They observed several features: the spectra exhibit a cluster of peaks; for an initial state near the metastable state, the peaks increase in amplitude and spacing with increasing frequency; for an initial state near the ground state, the peaks decrease in amplitude and spacing with increasing frequency; and in both cases, the peaks shift to lower frequencies and spread out with increasing confinement parameter. In earlier work, Pomponio et al.~\cite{Pomponio:2022} studied a different type of metastable oscillation. They performed an anti-confining quench and unexpectedly observed long-lived oscillations, interpreted as Bloch oscillations of the domain walls. The Fourier spectra of the correlation functions exhibited a regular peak structure.

Lagnese et al.~\cite{Lagnese:2023} accurately calculated the frequencies and amplitudes of the spectral peaks by restricting to the manifold of low-energy 0- and 1-meson states and examining the dynamics using perturbation theory. This followed Rutkevich~\cite{Rutkevich:1999} who used a similar approach to compute the tunneling probability in the decay of metastable excited states. We briefly summarize the model developed in these studies. It can be used more generally for any condensed matter system with confinement, and we thus use a more general notation in parts. In \Cref{subsec:quasiparticle:neutral_atoms}, we modify it to examine the dynamics in our neutral atom system.

The ferromagnetic Ising spin Hamiltonian is
\begin{equation}
H_I = - \sum_{i=1}^N ( \hat{\sigma}_{z,i} \hat{\sigma}_{z,i+1} + h_x \hat{\sigma}_{x,i} + h_z \hat{\sigma}_{z,i} ) \, ,
\label{eq:hamiltonian_ising}
\end{equation}
where $h_x$ and $h_z$ are the transverse and longitudinal fields. 
It can be written as $H_I = H_0 + U$. The first term $H_0 = - \sum_{i=1}^N ( \hat{\sigma}_{z,i} \hat{\sigma}_{z,i+1} + h_x \hat{\sigma}_{x,i} )$ is the transverse field model that has no confinement; the ground state is two-fold degenerate and domain walls propagate freely with dispersion relation $\omega(\theta) = 2 ( 1 - 2 h_x \cos\theta + h_x^2 )^{1/2}$ where $\theta$ is the quasimomentum. The second term $U = - \sum_{i=1}^N h_z \hat{\sigma}_{z,i}$ is the confinement term; it breaks the ground state degeneracy and introduces a linear interaction potential between domain walls. To use time-dependent perturbation theory effectively, it is necessary to rewrite the Hamiltonian as $H_I = \tilde{H}_0 + \tilde{U}$, where $U = U_0 + \tilde{U}$ and $\tilde{H}_0 = H_0 + U_0$~\cite{Rutkevich:1999,Lagnese:2023}. The term $U_0$ is the sum over the projections of $U$ onto subspaces with fixed meson number, and thus only couples states that have the same meson number. It strongly modifies the energy eigenvalues of $H_0$ due to the linear potential, and is thus included in the unperturbed Hamiltonian $\tilde{H}_0$. The remaining term $\tilde{U} = U - U_0$ then couples states with different meson number. It only weakly modifies the eigenvalues of $\tilde{H}_0$, and is thus treated as the perturbation to compute transitions between different meson sectors.

The perturbation conserves momentum $k$; at first order, it thus connects the metastable and ground states, which have $k=0$, to the 1-meson eigenstates with $k=0$, and only at higher orders to even superpositions of 1-meson eigenstates with degenerate energies and opposite momentum. To compute the 1-meson eigenstates with $k=0$, one can project the Schr\"{o}dinger equation for $\tilde{H}_0$ to the 1-meson sector, take the coordinate representation, and shift to the center of momentum frame $k=0$, obtaining~\cite{Rutkevich:1999,Lagnese:2023}
\begin{equation}
\sum_{n' > 0} T_{nn'} \phi_\ell (n') \pm U_0(n) \phi_\ell (n) = \frac{\varepsilon_\ell}{2} \phi_\ell (n) \, ,
\label{eq:1meson}
\end{equation}
where $n > 0$ is the domain size; $\ell > 0$ indexes the energy eigenstates $\phi_\ell (n)$ and eigenvalues $\varepsilon_\ell$; $T_{nn'}$ is the kinetic energy, or domain wall hopping, term arising from $H_0$, obtained from the Fourier transform of $\omega(\theta)$; and $U_0 (n) \propto \mathcal{E} n$ is the potential arising from $U_0$, written in terms of the confinement energy density $\mathcal{E} \propto h_z \mathcal{M}$, with $\mathcal{M} = (1 - h_x^2)^{1/8}$ the total magnetization of the ground states of $H_0$.

\subsection{Neutral Atom Setup}
\label{subsec:quasiparticle:neutral_atoms}

The model in \Cref{subsec:quasiparticle:spins} can be used in general condensed matter systems with confinement. We modify it for our neutral atom setup and examine the evolution of the magnetization. 

The Hamiltonian $H$ is given in \Cref{eq:hamiltonian}. The confinement term is $U = - \sum_j \Delta_{\mathrm{loc},j} \hat{n}_j$, where $\Delta_{\mathrm{loc},j}$ is the staggered local detuning field in \Cref{eq:local_detuning}. The Schr\"{o}dinger equation for 1-meson eigenstates with $k=0$ has the same form as \Cref{eq:1meson}, but the terms for the kinetic energy $T_{nn'}$ and potential $U_0 (n)$ are distinct to the Hamiltonian $H$. 
In lieu of directly computing $T_{nn'}$ and $U_0 (n)$, we examine them in relation to the classical limit in \Cref{sec:repulsion}. 

The evolved state $\ket{\psi(t)}$ can be approximated using standard techniques from time-dependent perturbation theory. The initial state $\ket{\psi(0)}$ is approximately the desired metastable or ground state of $\tilde{H}_0$, which can be denoted $\ket{0_v}$ with eigenvalue $\varepsilon_v$ where $v = m,g$ labels the state type. To first order in perturbation theory, as examined in Ref.~\cite{Lagnese:2023}, the evolved state is
\begin{equation}
\ket{\psi(t)} \approx e^{-i \varepsilon_v t} \ket{0_v} + \sum_\ell c_\ell (t) e^{-i \varepsilon_\ell t} \ket{\phi_\ell} ,
\label{eq:psi_of_t}
\end{equation}
where $c_\ell (t) = d_\ell (t) \langle \phi_\ell | \tilde{U} | 0_v \rangle$ are the first order coefficients, with $d_\ell (t) = (e^{i \tilde{\omega}_{\ell v} t} - 1)/(i \tilde{\omega}_{\ell v})$, using the frequencies $\tilde{\omega}_{\ell v} = \varepsilon_\ell - \varepsilon_v$. The expression used the result $c_{v'} (t) \propto \langle 0_{v'} | \tilde{U} | 0_v \rangle = 0$ since $\tilde{U} = U - U_0$ only connects states with different meson numbers. Furthermore, the matrix element satisfies $\langle \phi_\ell | \tilde{U} | 0_v \rangle = \langle \phi_\ell | U | 0_v \rangle$ since $U_0$ only connects states with the same meson number.

We characterize the evolved state by the magnetization $M = \langle \hat{M} \rangle$. To first order in perturbation theory, it is
\begin{equation}
\begin{aligned}
\langle \hat{M} \rangle \approx \,\, & \langle 0_v | \hat{M} | 0_v \rangle \\
& + \sum_\ell 2 \operatorname{Re} \left\lbrace d_\ell^* (t) \langle \phi_\ell | \tilde{U} | 0_v \rangle \langle 0_v | \hat{M} | \phi_\ell \rangle \right\rbrace ,
\end{aligned}
\label{eq:M_of_t}
\end{equation}
where $d_\ell^* (t) = (1 - e^{-i \tilde{\omega}_{\ell v} t})/(i \tilde{\omega}_{\ell v})$. The major peaks in the Fourier spectrum of $M$ can be understood from this expression. The frequency of a major peak equals a frequency $\tilde{\omega}_{\ell v} = \varepsilon_\ell - \varepsilon_v$ between one of the meson states $\ket{\phi_\ell}$ and the state $\ket{0_v}$. The energies $\varepsilon_v$ and $\varepsilon_\ell$ are the eigenvalues of the unperturbed Hamiltonian $\tilde{H}_0$; however, since the perturbation is weak, these are approximately equal to the eigenvalues of the exact Hamiltonian $H$. The amplitude of the peak depends on the two matrix elements $\langle \phi_\ell | \tilde{U} | 0_v \rangle$ and $\langle 0_v | \hat{M} | \phi_\ell \rangle$. The Fourier spectrum of $M$ can thus probe interference between the desired state and proximate 1-meson states, connected by $\tilde{U}$ and $\hat{M}$, which can potentially identify it. 

In rough terms, the Hamming distance of the state $\ket{\phi_\ell}$ from the state $\ket{0_v}$, dominated by the $Z_2$ product state $\ket{10 \hdots 10}$, is inversely correlated with the size of the matrix elements $\langle \phi_\ell | \tilde{U} | 0_v \rangle$ and $\langle 0_v | \hat{M} | \phi_\ell \rangle$. This can be explained as follows. The energies of the 1-meson eigenstates are inversely correlated with the domain size for a metastable initial state, and a subset are correlated for a ground initial state (\Cref{sec:repulsion} examines these trends in detail). As the domain size increases, the non-zero terms in the matrix elements arise from higher order domain wall hoppings. In lieu of directly computing the matrix elements, we use the Hamming distance to estimate trends. 

We note that Lagnese et al.~\cite{Lagnese:2023} also performed a supplementary study on an antiferromagnetic 1D neutral atom chain. They used a 24-atom chain, restricted to nearest-neighbor interactions, with the corresponding blockade subspace, and examined a specific set of parameters. They interpreted the spectra in analogy with their results for the spin chain. However, from our extensive studies with a more general model that includes long-range interactions, we identify spectral features that are distinct to these systems (\Cref{sec:numerical}) and explain them only from the specific microphysics in the Hamiltonian (\Cref{sec:repulsion}). 
\section{Numerical Study}
\label{sec:numerical}

\begin{figure}
\includegraphics[width=0.48\textwidth]{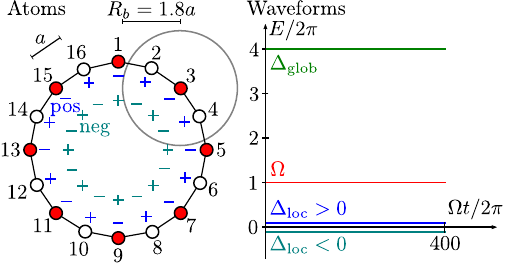}
\caption{The numerical setup to study metastable and ground state long-lived oscillations. The left section shows the atom configuration. The grey circle shows the blockade radius. The initial state is set to the ground state of $H_\mathrm{prep} = H(\beta_\mathrm{prep})$ with $\beta_\mathrm{prep} = -10^{-3}$; it is dominated by the $Z_2$ product state $\ket{10 \hdots 10}$, represented by the color pattern for the atoms. The state then evolves under $H$ with a different $\beta$. The right section shows the waveform sequence. The waveforms are all held constant. The local detuning field for a small positive value $\beta \propto \Delta_\mathrm{loc} > 0$ is shown by the blue outer $+/-$ symbols on the left and the labeled waveform on the right, producing an initial state close to the metastable state of $H$. The field for a small negative value $\beta \propto \Delta_\mathrm{loc} < 0$ is shown by the teal inner $+/-$ symbols on the left and the labeled waveform on the right, producing an initial state close to the ground state of $H$.}
\label{fig:setup_oscillations}
\end{figure}

We study the long-lived oscillations of metastable and ground states in 1D neutral atom systems, using the setup shown in \Cref{fig:setup_oscillations}. The atoms lie in a ring with $n_s$ sites, where we choose $n_s = 16$. In particular, we use a staggered local detuning field to achieve confinement, yielding a confinement energy density $\mathcal{E} \propto \Delta_\mathrm{loc}$. We set the initial state $\ket{\psi(0)}$ to be the ground state of $H_\mathrm{prep} = H(\beta_\mathrm{prep})$, obtained from exact diagonalization, where $H$ is the Rydberg Hamiltonian in \Cref{eq:hamiltonian} and $\beta_\mathrm{prep} = -10^{-3}$. It is dominated by the $Z_2$ product state $\ket{10 \hdots 10}$, represented by the colors in the atom chain in \Cref{fig:setup_oscillations}. We evolve the state under $H$ with a different local detuning parameter $\beta$, which has a small constant amplitude $| \beta | \ll 1$. In effect, we perform a quench $\delta \beta = \beta - \beta_\mathrm{prep}$. The initial state is then close to the metastable state of $H$ for $\beta \propto \Delta_\mathrm{loc} > 0$, shown by the blue outer $+/-$ pattern and waveform in \Cref{fig:setup_oscillations}; it is close to the ground state of $H$ for $\beta \propto \Delta_\mathrm{loc} < 0$, shown by the teal inner $+/-$ pattern and waveform. We note that in our procedure here, we hold the initial state fixed and modify the spatial pattern to obtain proximity to the metastable and ground states; this is the reverse of the schematic in \Cref{fig:overview}, where the spatial pattern is held fixed and the different states show the proximity.

The parameter requirements for the oscillations regime are given by two broad sets of conditions. The first are $V_2 \ll \Omega < \Delta_\mathrm{glob} \ll V_1$, so that the system is deep in the $Z_2$ phase and the large Rydberg blockade heavily suppresses nucleation. The second is $| \Delta_\mathrm{loc} | n_s \ll \Delta_\mathrm{glob}$, so that critical bubbles of size $\ell \sim \Delta_\mathrm{glob} / \Delta_\mathrm{loc} < n_s$ are suppressed. We thus center our study around the following parameter values: a large nearest-neighbor blockade spacing $R_b / a = 1.8$ to achieve a nearest-neighbor interaction strength $V_1$ that is much greater than $\Omega$, $\Delta_\mathrm{glob}$, and $| \Delta_\mathrm{loc} |$; a modest global detuning parameter $\alpha  = 4.0$; and a small constant local detuning amplitude $| \beta | \in [0.002, 0.15] \ll 1$ to achieve the subcritical condition $| \Delta_\mathrm{loc} | n_s \ll \Delta_\mathrm{glob}$. 

\begin{figure*}
\subfloat[$\beta = 0.01$]{\label{fig:M_vs_t_beta0.01}\includegraphics[width=0.48\textwidth]{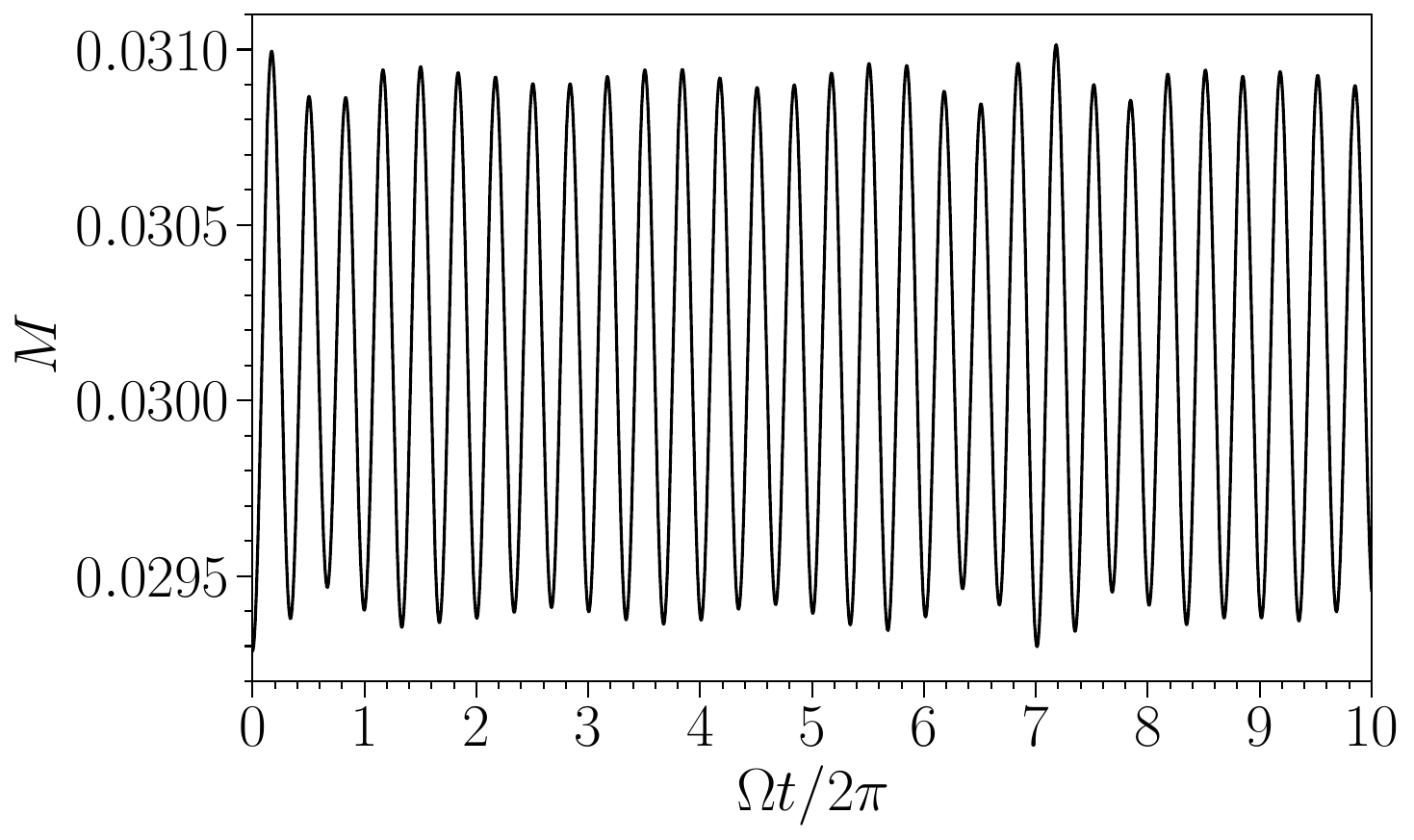}}\hfill
\subfloat[$\beta = 0.01$]{\label{fig:logMw_vs_w_beta0.01}\includegraphics[width=0.47\textwidth]{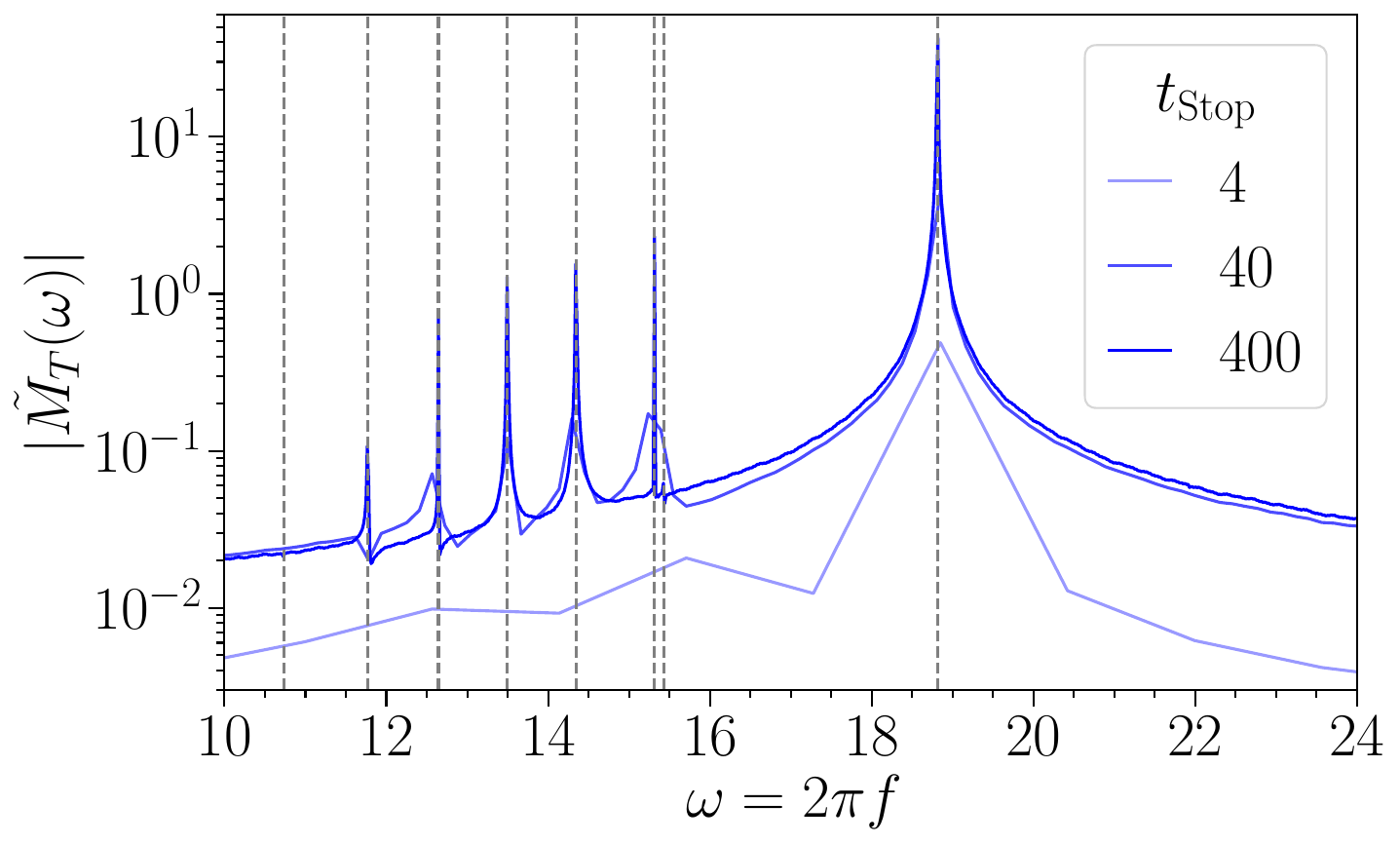}}\\
\subfloat[$\beta = \pm 0.01$]{\label{fig:eigenvalues_beta0.01}\includegraphics[width=0.526\textwidth]{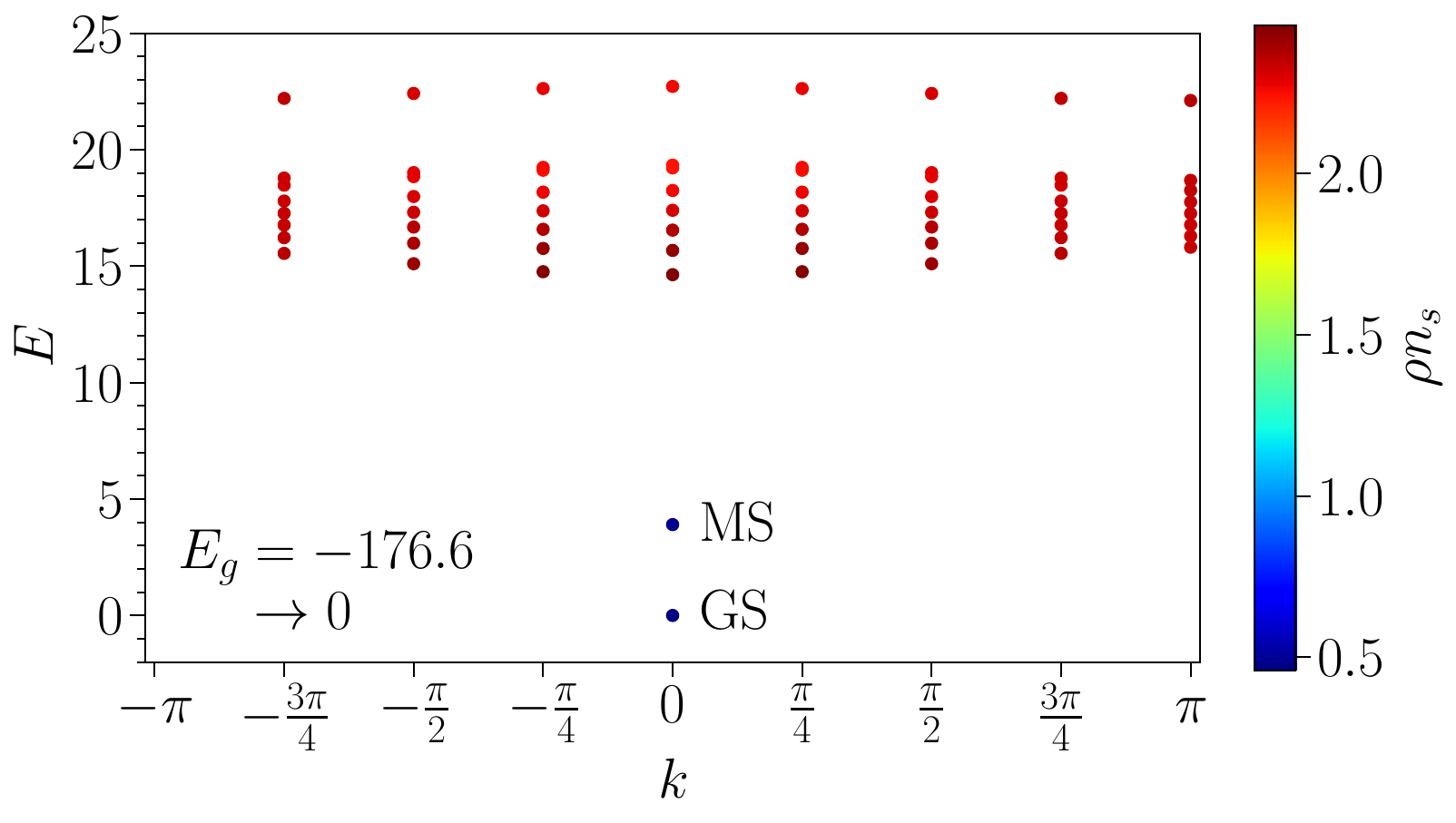}}\hfill
\subfloat[$\beta = -0.01$]{\label{fig:logMw_vs_w_beta-0.01}\includegraphics[width=0.47\textwidth]{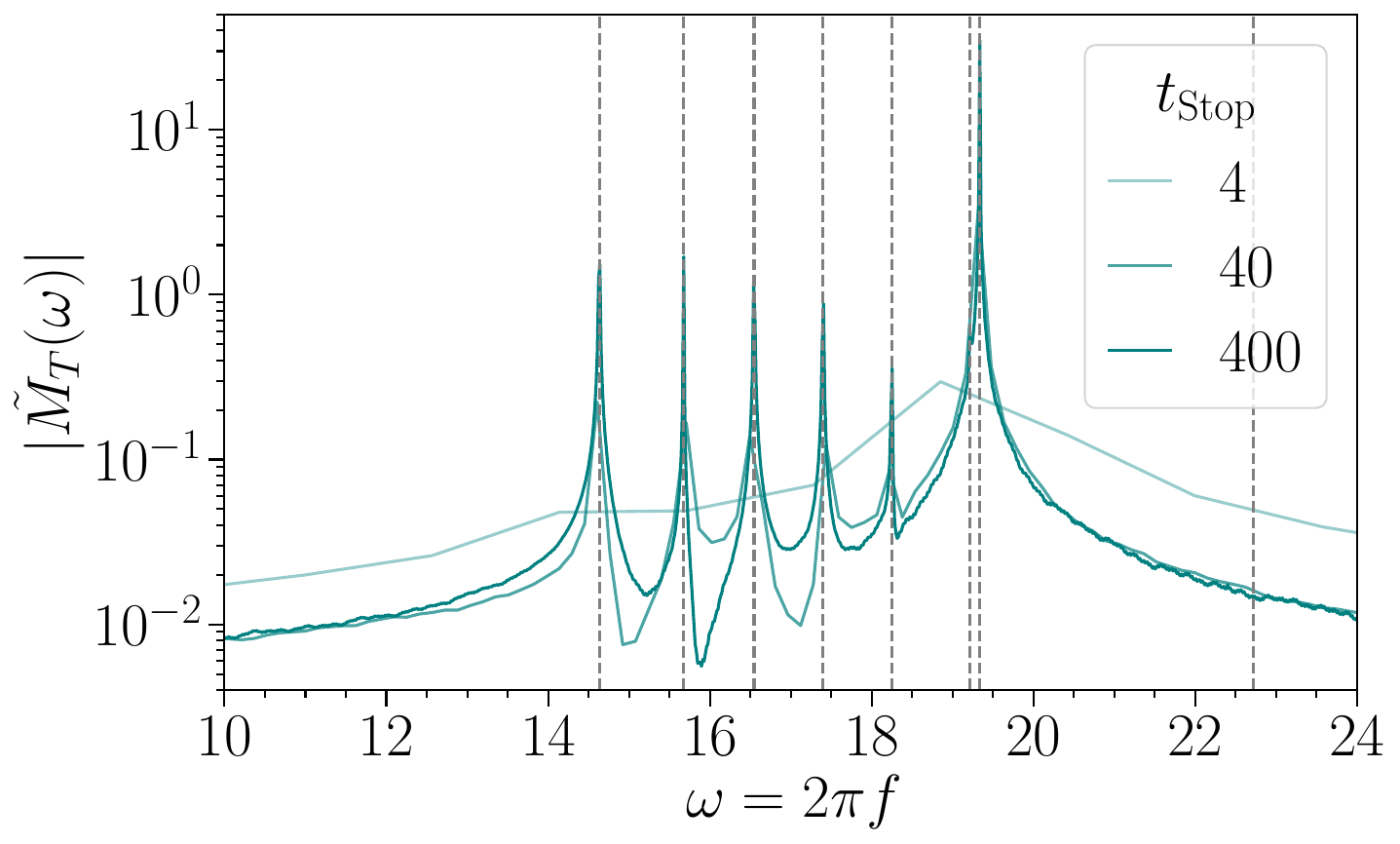}}
\caption{The oscillation properties of the magnetization $M$ for $\beta = \pm 0.01$. (a) The magnetization $M$ vs (dimensionless) time $\Omega t / 2\pi$ for $\beta = 0.01$. 
(b) The Fourier magnitude $|\tilde{M}_T(\omega)|$ of the transformed magnetization $M_T$, defined in \Cref{eq:M_T}, for $\beta = 0.01$, which yields an initial state close to the metastable state. The curves with different opacities show results for $M_T$ evolved to different stop times $t_\mathrm{Stop}$. (c) The simultaneous eigenstates of $H$ and $T_2$. The eigenstates are identical for $\beta = \pm 0.01$, by symmetry. The ground state energy is set to zero, $E_g \rightarrow 0$, as indicated in the bottom left. The color bar shows the domain wall density $\rho n_s$; the blue points show the metastable state (MS) and ground state (GS), and the red points show the $1$-meson eigenstates. (d) The Fourier magnitude $|\tilde{M}_T(\omega)|$ for $\beta = -0.01$, which yields an initial state close to the ground state. In (b) and (d), the vertical grey lines show the frequencies $\omega_{\ell v} = E_\ell - E_v$, where $E_\ell$ for $\ell = 1, \hdots, 8$ are the energy eigenvalues of the 1-meson eigenstates with momentum $k=0$ and $E_v$ is the energy eigenvalue of the metastable or ground state $v = m, g$. The Fourier spectra for $\beta = \pm 0.01$ both have a similar dominant peak but different clusters of secondary peaks, which can be used to distinguish between the metastable and ground states, and understood from our analysis in \Cref{sec:repulsion}.}
\label{fig:oscillation_properties}
\end{figure*}

\begin{figure*}
\subfloat[$\beta = 0.02$]{\label{fig:logMw_vs_w_beta0.02}\includegraphics[width=0.32\textwidth]{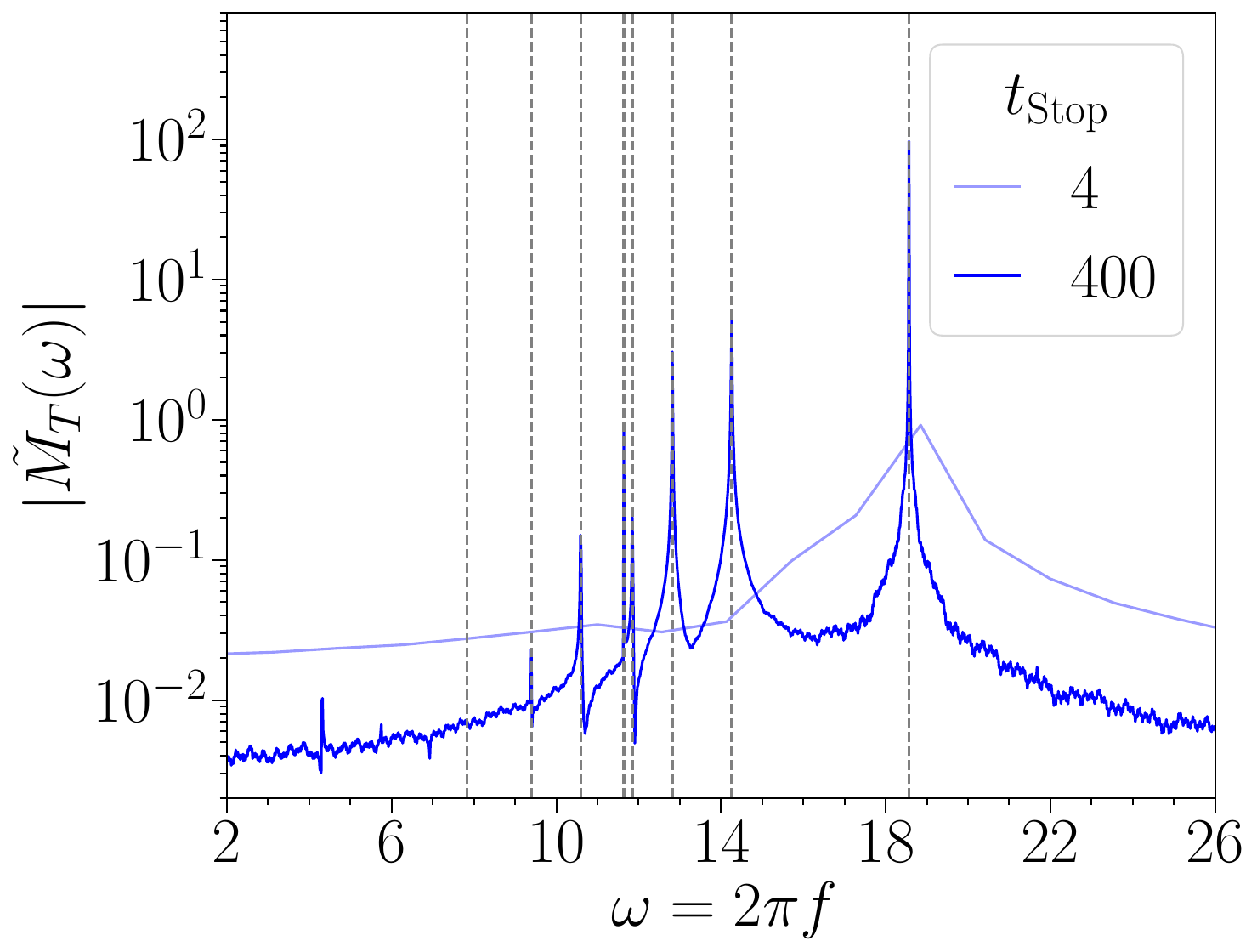}}\hfill
\subfloat[$\beta = -0.02$]{\label{fig:logMw_vs_w_beta-0.02}\includegraphics[width=0.32\textwidth]{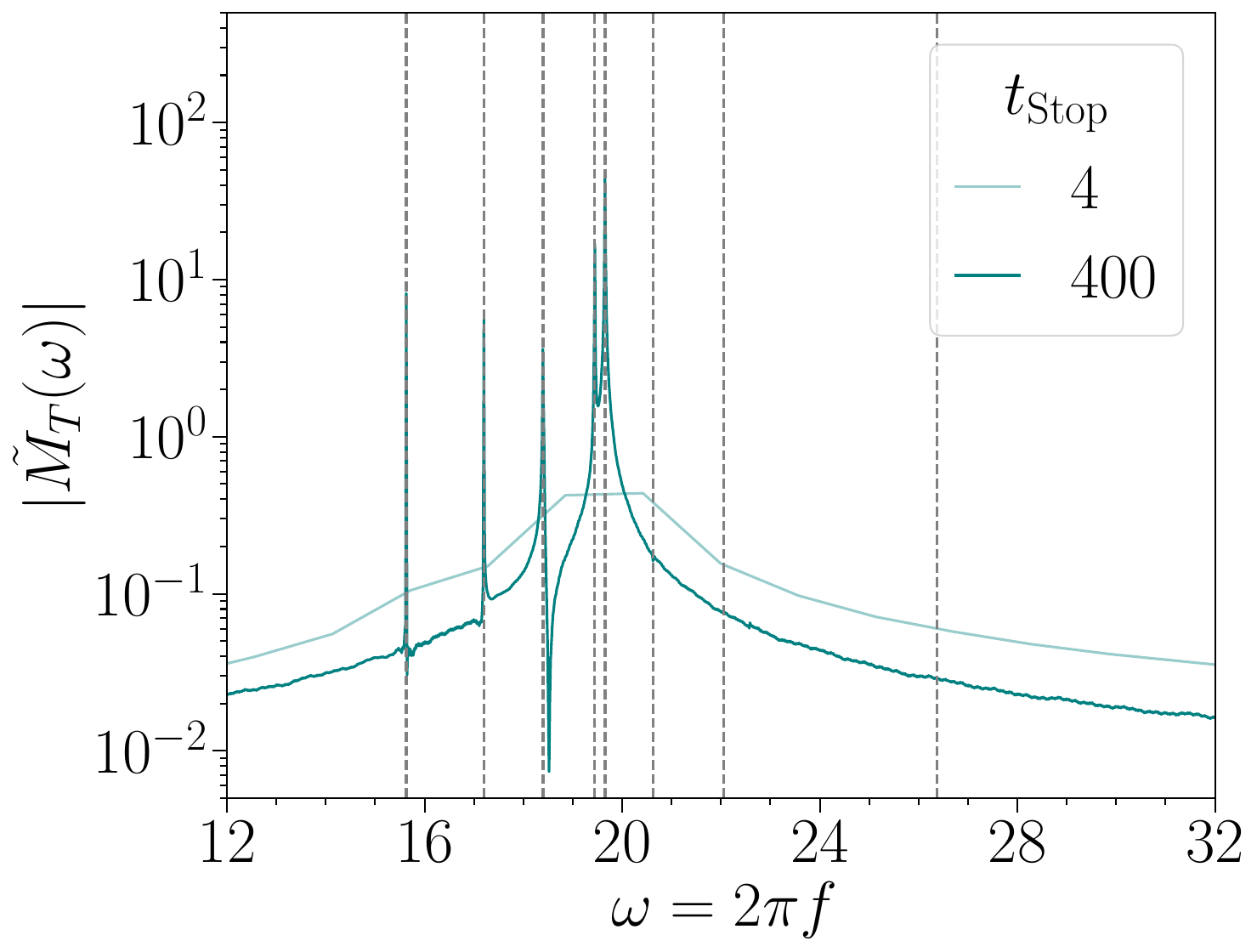}}\hfill
\subfloat[$\beta = \pm 0.02$]{\label{fig:eigenvalues_beta0.02}\includegraphics[width=0.35\textwidth]{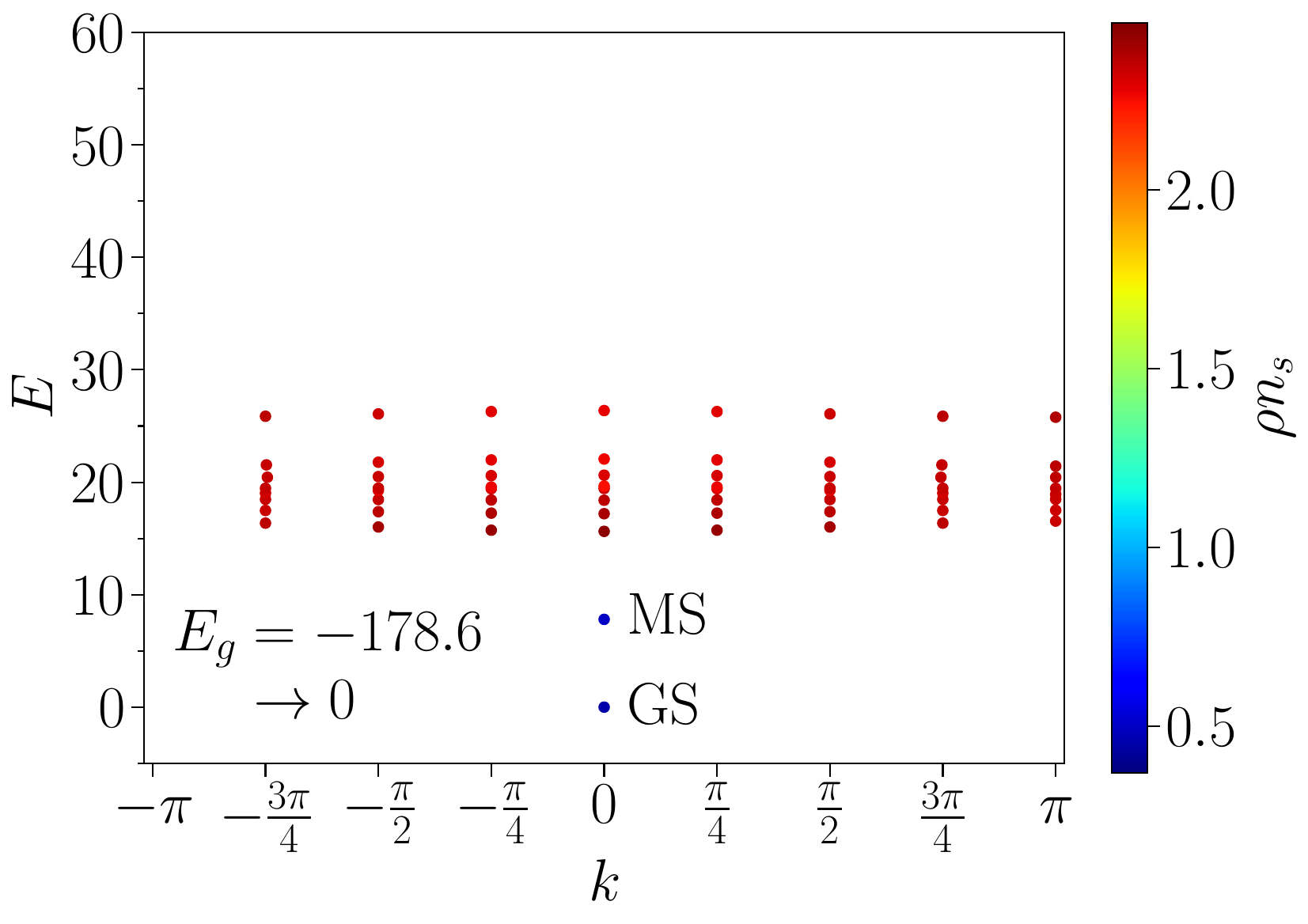}}\\
\subfloat[$\beta = 0.06$]{\label{fig:logMw_vs_w_beta0.06}\includegraphics[width=0.32\textwidth]{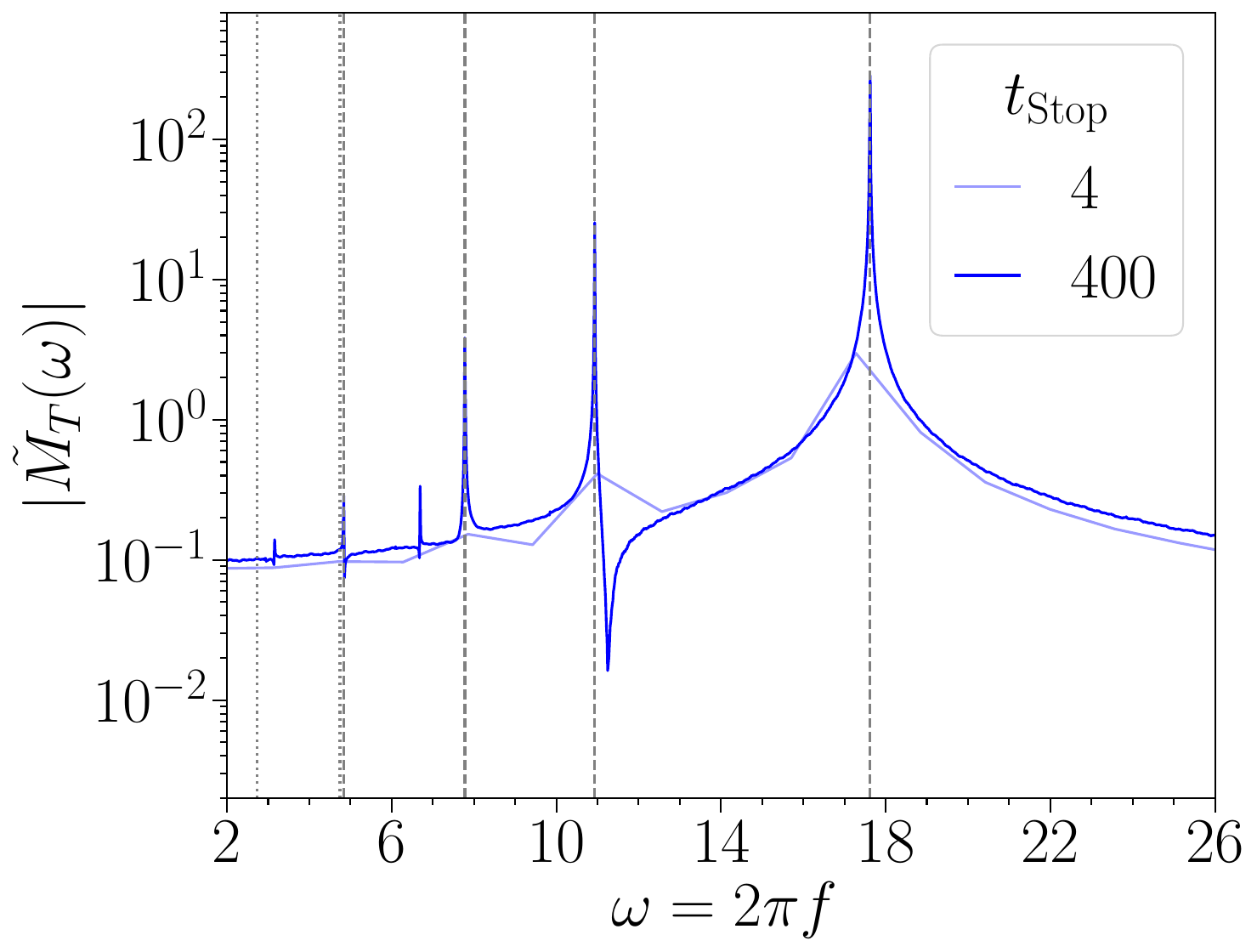}}\hfill
\subfloat[$\beta = -0.06$]{\label{fig:logMw_vs_w_beta-0.06}\includegraphics[width=0.32\textwidth]{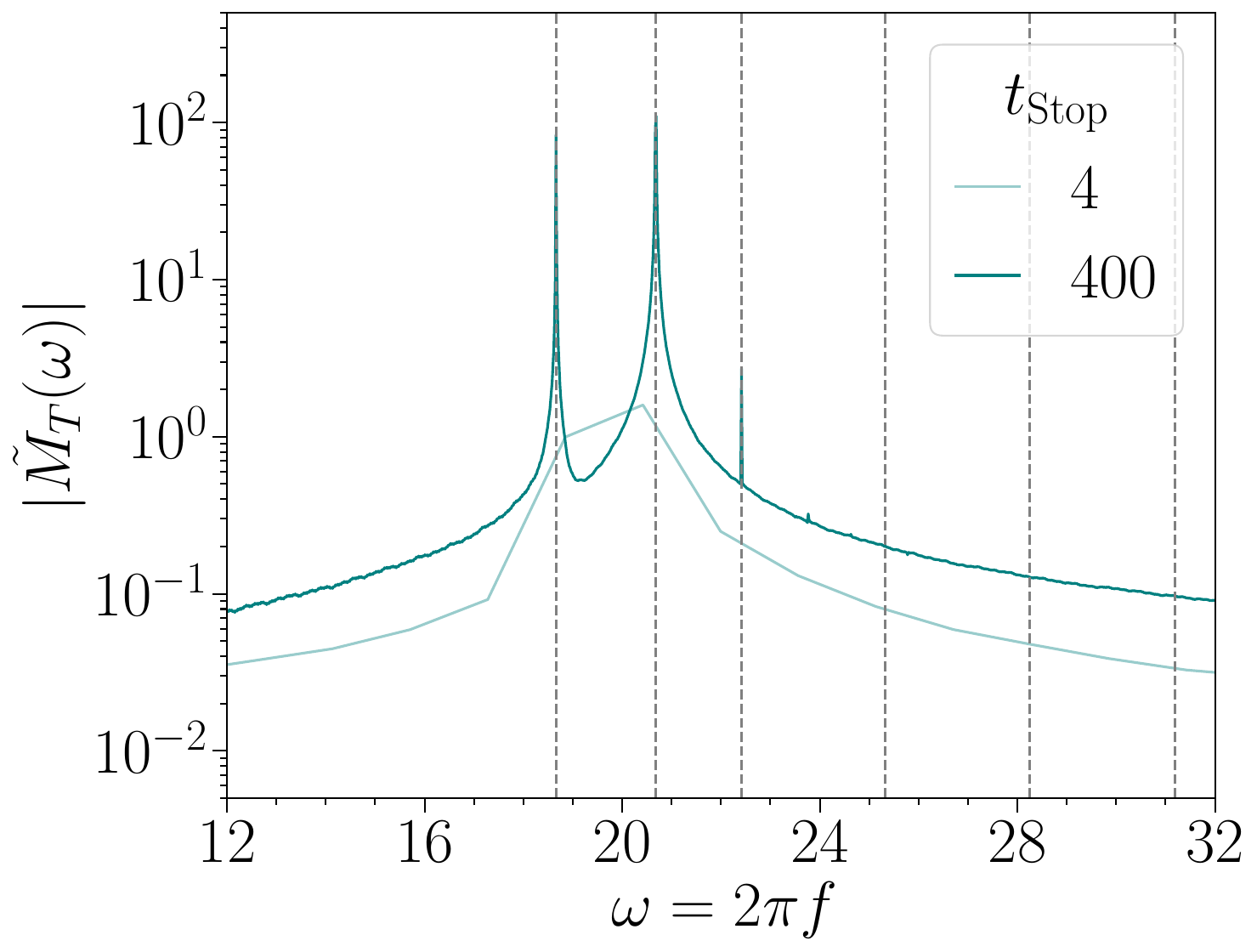}}\hfill
\subfloat[$\beta = \pm 0.06$]{\label{fig:eigenvalues_beta0.06}\includegraphics[width=0.35\textwidth]{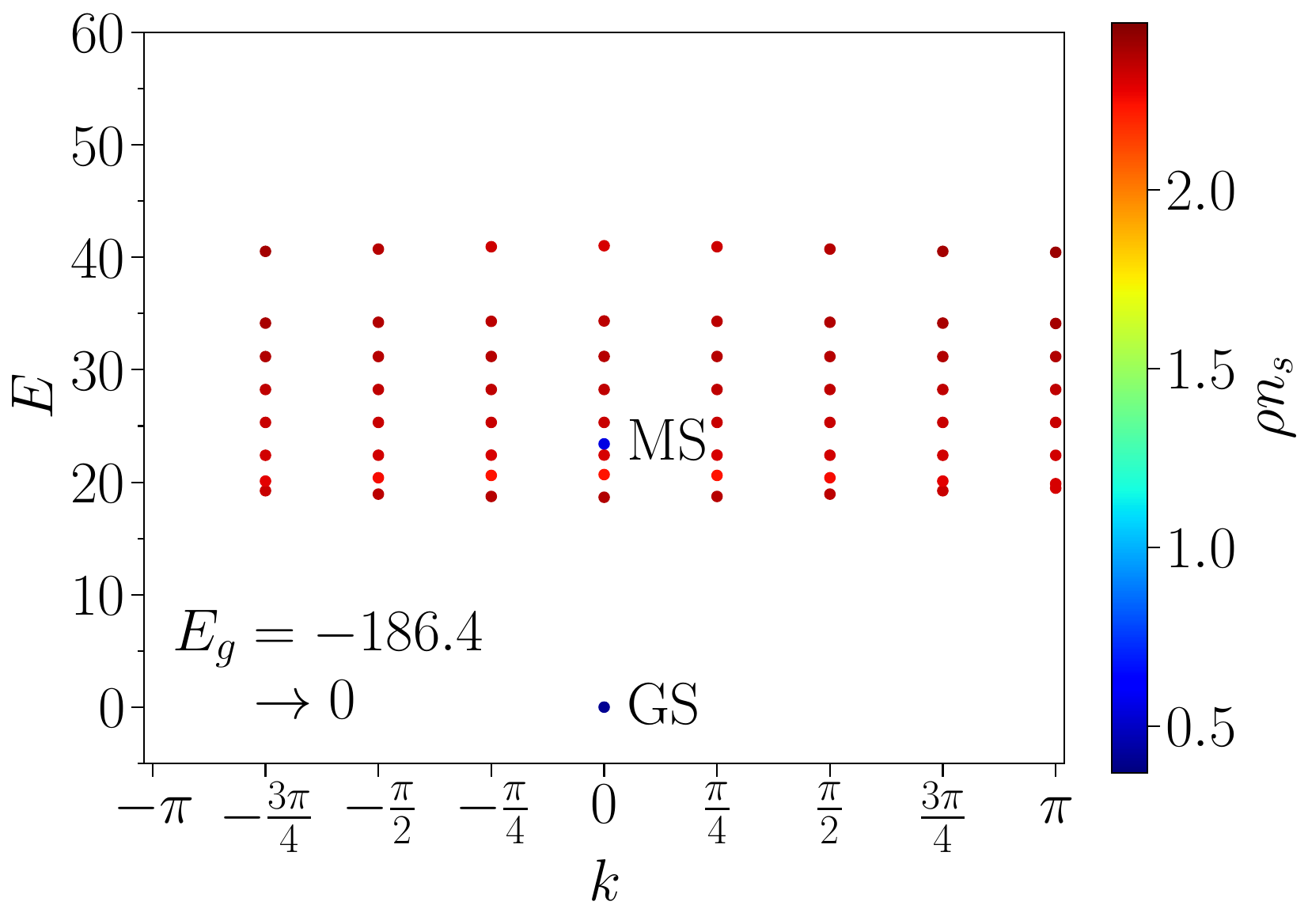}}\\
\subfloat[$\beta = 0.10$]{\label{fig:logMw_vs_w_beta0.10}\includegraphics[width=0.32\textwidth]{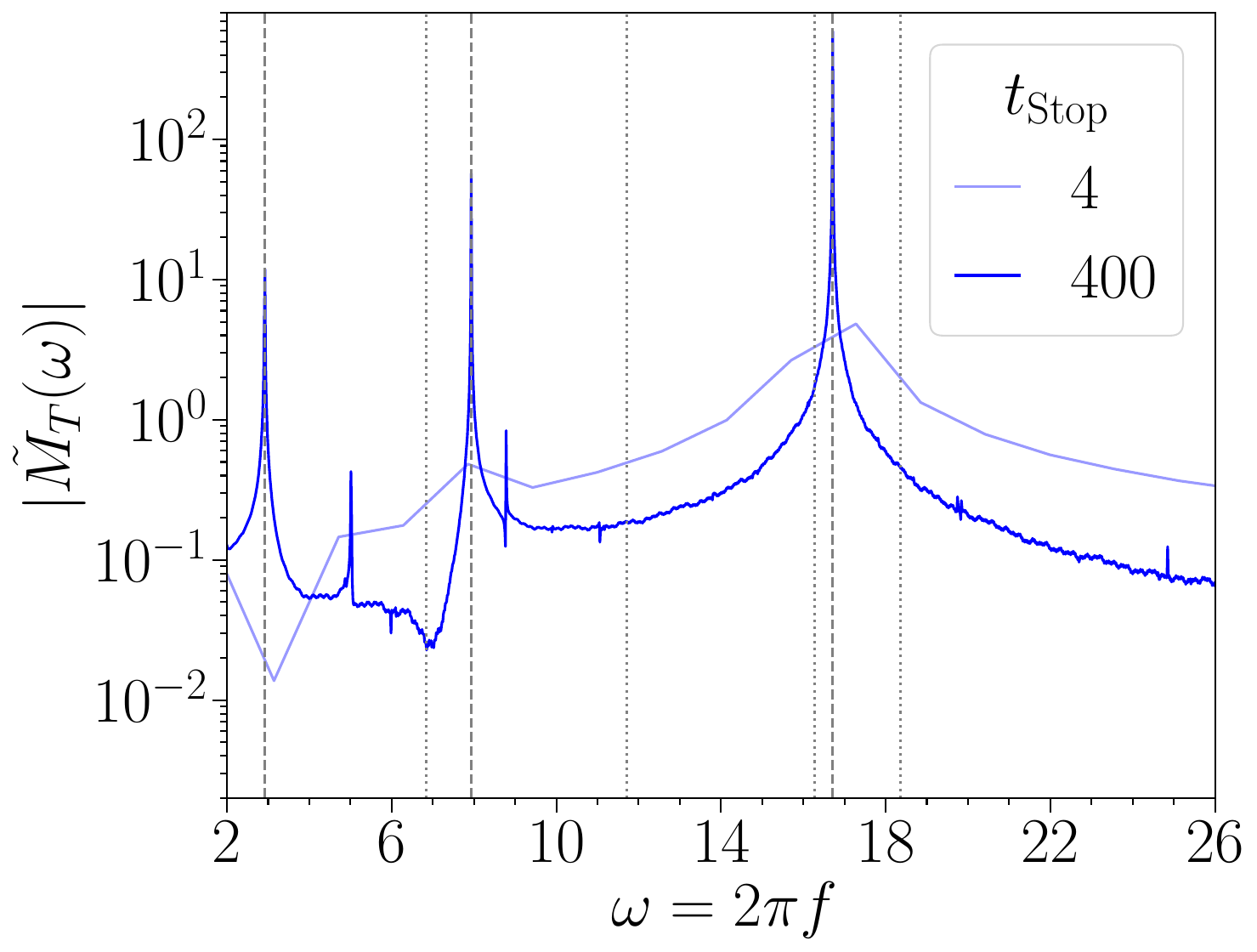}}\hfill
\subfloat[$\beta = -0.10$]{\label{fig:logMw_vs_w_beta-0.10}\includegraphics[width=0.32\textwidth]{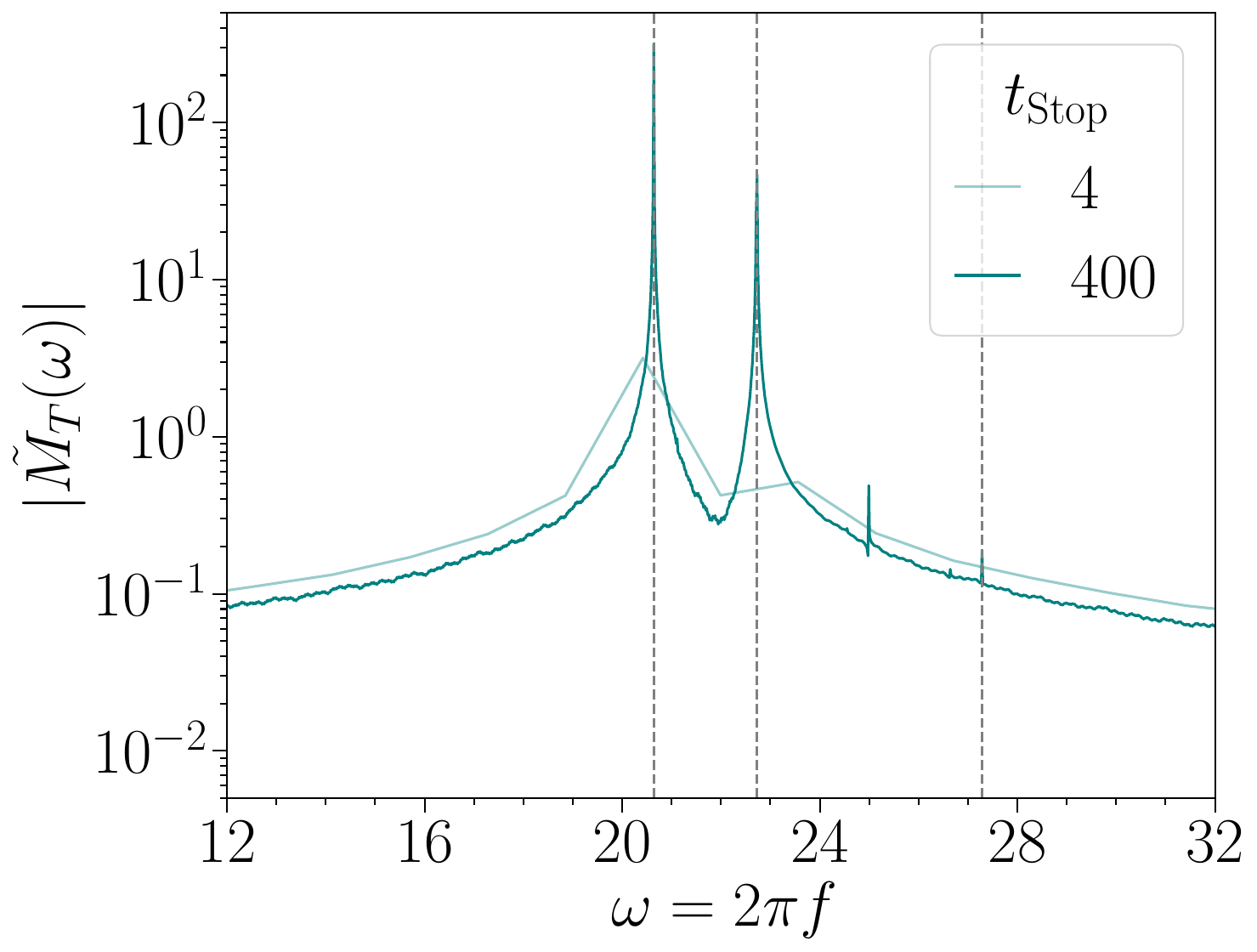}}\hfill
\subfloat[$\beta = \pm 0.10$]{\label{fig:eigenvalues_beta0.10}\includegraphics[width=0.35\textwidth]{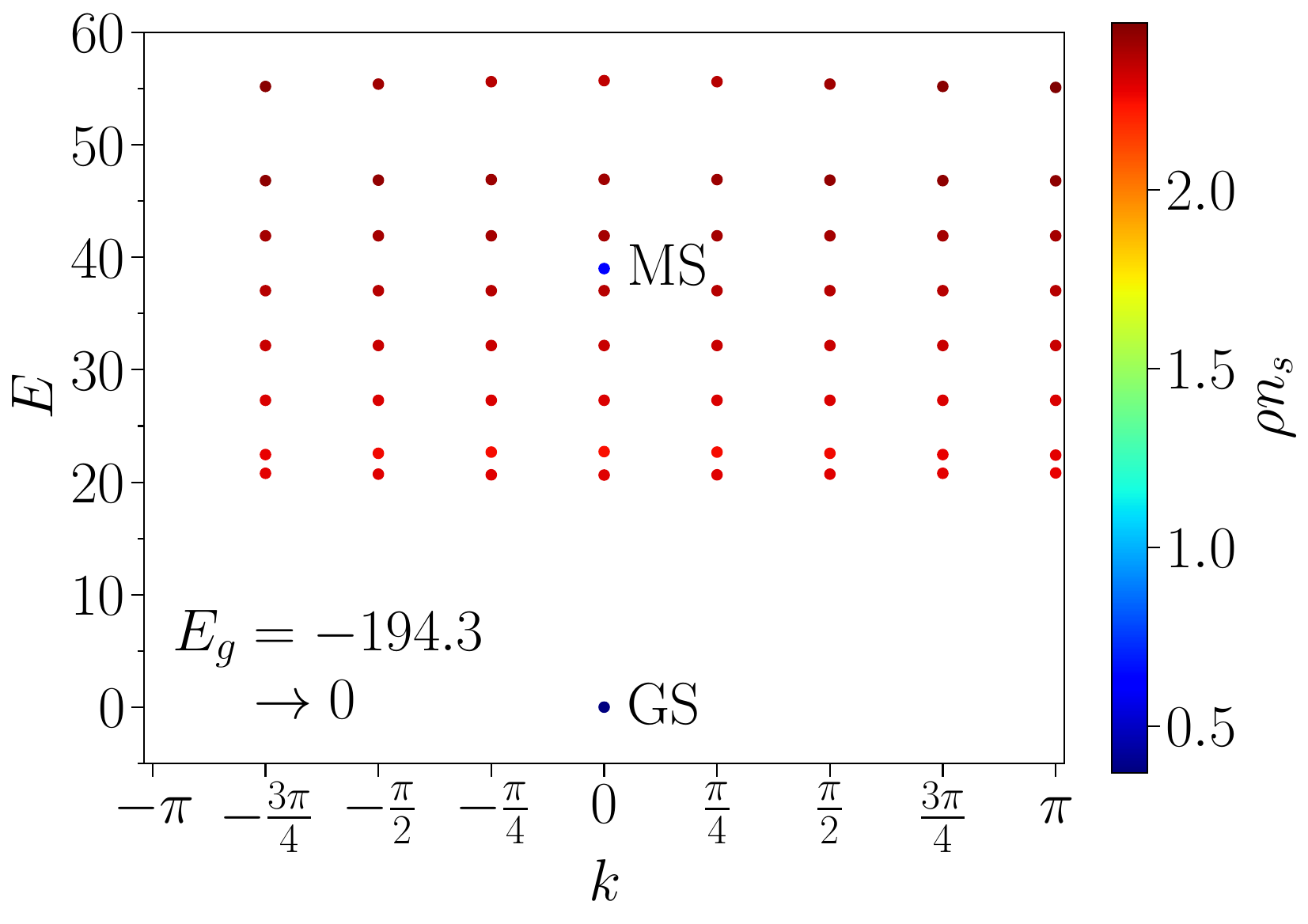}}
\caption{The Fourier and eigenvalue spectra for $\beta = \pm 0.02, 0.06, 0.10$. Left column: The Fourier magnitude $|\tilde{M}_T(\omega)|$ for several values of $\beta > 0$, which yield an initial state close to the metastable state. The curve opacities and vertical grey lines are described in \Cref{fig:oscillation_properties}. The dashed lines show frequencies $\omega_{\ell v} > 0$ and the dotted lines show frequencies $\omega_{\ell v} < 0$. Middle column: The Fourier magnitude $|\tilde{M}_T(\omega)|$ for several values of $\beta < 0$, which yield an initial state close to the ground state. Right column: The simultaneous eigenstates of $H$ and $T_2$. The conventions and color bar are described in \Cref{fig:oscillation_properties}. In the Fourier spectra for $\beta > 0$ and $\beta < 0$, the secondary peaks shift in opposite directions, which can be used to distinguish between the metastable and ground states, and understood from the trends in the eigenvalue spectra and our analysis in \Cref{sec:repulsion}.}
\label{fig:spectra_beta_large}
\end{figure*}

\subsection{Metastable case: \texorpdfstring{$\beta > 0$}{B > 0}}
\label{subsec:numerical:metastable}

We first consider an initial state close to the metastable state. We center our discussion around the case $\beta = 0.01$, since it captures many important features and serves as an effective base case to describe trends with increasing $\beta$. \Cref{fig:M_vs_t_beta0.01} shows the magnetization, which exhibits long-lived oscillations with several dominant frequencies. \Cref{fig:logMw_vs_w_beta0.01} shows the Fourier spectrum of the transformed magnetization $M_T$ when it is evolved to different stop times $t_\mathrm{Stop}$. Naturally, a time-domain signal with a longer duration produces a frequency-domain signal with a higher frequency resolution. The stop times each increase by an order of magnitude to capture the improving resolution, and the lowest value $t_\mathrm{Stop} = 4.0$ is closest to current experimental constraints for $\Omega / 2\pi = 1.0$ MHz (\Cref{sec:experiments}). The highest-resolution spectrum has a large, wide, dominant peak at $\omega \approx 19$ and a cluster of smaller, narrower, evenly-spaced secondary peaks at lower frequencies $\omega \in [11,16]$. The secondary peaks increase in amplitude with increasing frequency. Though not shown, the N\'{e}el OP has a similar Fourier spectrum to the magnetization. 

In \Cref{sec:quasiparticle}, we outlined a perturbative formalism to interpret the oscillation spectra. As we presented in \Cref{eq:1meson} and the surrounding text, the peaks in the magnetization spectra depend on the energy eigenvalues and eigenvectors of the unperturbed Hamiltonian $\tilde{H}_0$. However, in our analysis in this section, we replace these with the exact eigenvalues and eigenvectors of $H$ since the perturbation $\tilde{U}$ is weak, and denote them by the base variable $E$. 

\Cref{fig:eigenvalues_beta0.01} shows the simultaneous eigenstates of $H$ and $T_2$, with energy eigenvalues $E$ and translation eigenvalues $e^{ik}$ (equivalently, momenta $k$). We computed the eigenvalues by exact diagonalization with matrix balancing, a similarity transformation that improves the numerical stability~\cite{Parlett:1969,James:2014}. The figure shows only the low-energy $0$- and $1$-meson eigenstates, those selected with energy $E/\Omega \leq 500$ and domain wall density $\rho n_s \lesssim 3$. The blue points show the metastable (MS) and ground (GS) states; these have $k=0$ and $\rho n_s \lesssim 1$, i.e. approximately $\rho n_s \approx 0$ domain walls. We label the energy eigenvalues of the metastable and ground states as $E_m$ and $E_g$, respectively. The energy separation $\omega_{mg} = E_m - E_g$ between the two states is proportional to the local detuning parameter, $\omega_{mg} \propto \Delta_\mathrm{loc}$, as required since $\omega_{mg} \propto \mathcal{E}$. The red points show the 1-meson eigenstates; these have $1.5 \lesssim \rho n_s \lesssim 2.5$, i.e. approximately $\rho n_s \approx 2$ domain walls. 

The band structure can be explained as follows. The parameters lie in the range $| \Delta_\mathrm{loc} | n_s \ll \Delta_\mathrm{glob}$ and $V_2 \ll \Delta_\mathrm{glob}$, which has several consequences. The generation of two domain walls comes with an energy increase $\Delta_\mathrm{glob}$, which is much greater than the energy decrease possible from any domain size growth. The excited states thus split into separated manifolds, where the states in a single manifold approximately correspond to a single number $N$ of mesons. The Hamiltonian approximately conserves the meson number $N$. In the low-energy sector, manifolds with higher energy have larger $N$. In particular, the metastable and ground states approximately correspond to 0-mesons, and the manifold of first excited states correspond to 1-mesons. The confinement field splits the manifold of 1-meson eigenstates into a tower of eight meson bands, approximately corresponding to the $n_s / 2 = 8$ odd size domains $n = 1, 3, \hdots n_s-1$, as we discuss below in \Cref{sec:repulsion}. Each band splits into $n_s / 2 = 8$ momenta $k$, since $T_2$ is a 2-site translation operator. 

In agreement with our model in \Cref{subsec:quasiparticle:neutral_atoms}, we find that the peaks arise from interference between the 1-meson eigenstates with $k=0$ and the metastable state. In other words, the peaks are located at the frequencies $\omega_{\ell m} = E_\ell - E_m$, where $E_\ell$ for $\ell = 1, \hdots, 8$ are the energies of the $1$-meson eigenstates with momentum $k=0$, arranged in the order $E_1 < \hdots < E_8$. The quench $\delta \beta = \beta - \beta_\mathrm{prep}$ conserves momentum, so it can only excite 1-meson eigenstates with the same momentum as the metastable state, $k=0$. The quench can also excite a superposition of degenerate 1-meson eigenstates with momenta $\pm k$, but this is a higher order process. The observed frequencies support the 1-meson model in \Cref{eq:1meson}. 

The frequencies $\omega_{\ell m}$ are shown by the vertical grey lines in \Cref{fig:logMw_vs_w_beta0.01}. The dominant peak is located at the frequency $\omega_{8m}$. We can explain this as follows: out of all the 1-meson states $\ket{E_\ell}$, the dominant basis elements in the state $\ket{E_8}$ have the smallest Hamming distance from the $Z_2$ product state $\ket{10 \hdots 10}$ that dominates $\ket{E_m}$, producing the largest matrix elements in \Cref{eq:M_of_t}. The cluster of secondary peaks are located at the frequencies $\omega_{\ell m}$ for $\ell = 1, \hdots, 7$. They generally start with large amplitude at $\omega_{7m}$ and diminish in amplitude as they approach $\omega_{1m}$, eventually vanishing altogether. Again, we can explain this trend with the Hamming distance: as $\ell$ decreases from $7$ to $1$, the dominant basis elements in the states $\ket{E_\ell}$ increase in Hamming distance from the $Z_2$ product state and overlap with the other basis states of $\ket{E_m}$, which themselves decrease in amplitude with increasing Hamming distance from the $Z_2$ product state, so that the corresponding matrix elements decrease. 

We observe several trends in the spectrum as we increase $\beta > 0$. These are captured in Figures \ref{fig:logMw_vs_w_beta0.02}, \ref{fig:logMw_vs_w_beta0.06}, \ref{fig:logMw_vs_w_beta0.10}. The dominant peak shifts modestly to the left. The cluster of secondary peaks shift more rapidly to the left and spread out. The trends in the energy eigenvalue distribution provide a complementary picture, as shown in Figures \ref{fig:eigenvalues_beta0.02}, \ref{fig:eigenvalues_beta0.06}, \ref{fig:eigenvalues_beta0.10}. The metastable and ground state energies separate, with the metastable energy moving upwards towards, and eventually through, the 1-meson energies. The 1-meson energies spread uniformly. The separation between $E_m$ and $E_8$ remains roughly constant, consistent with the dominant peak at $\omega_{8m}$ remaining at roughly the same location. The separation between $E_m$ and the remaining energies $E_\ell$ for $\ell = 1, \hdots, 7$ decreases, consistent with the secondary peaks at $\omega_{\ell m}$ shifting to the left.

\subsection{Ground case: \texorpdfstring{$\beta < 0$}{B < 0}}
\label{subsec:numerical:ground}

We next consider an initial state close to the ground state, beginning with the case $\beta = -0.01$. \Cref{fig:logMw_vs_w_beta-0.01} shows the Fourier spectrum of $M_T$. The highest-resolution spectrum for $\beta = -0.01$ is similar to the one for $\beta = 0.01$. However, the secondary peaks are located at higher frequencies $\omega \in [15, 18]$ and they generally decrease in amplitude with increasing frequency. The simultaneous eigenstates of $H$ and $T_2$ for $\beta = -0.01$ are identical to those for $\beta = 0.01$, since the Hamiltonian is invariant under the simultaneous transformation $\beta \rightarrow -\beta$ and $H \rightarrow T_1 H T_1^\dagger$, where $T_1$ is the 1-site translation operator. 

The peaks for the case $\beta = -0.01$ are located at the frequencies $\omega_{\ell g} = E_\ell - E_g$, shown by the vertical grey lines in \Cref{fig:logMw_vs_w_beta-0.01}. The dominant peak is located at a frequency $\omega_{\sigma g}$ for a particular index $\sigma$, which can be explained as follows: out of all the states $\ket{E_\ell}$, the dominant basis elements of the state $\ket{E_\sigma}$ have the smallest Hamming distance from the $Z_2$ product state $\ket{10 \hdots 10}$ that dominates $\ket{E_g}$, producing the largest matrix elements in \Cref{eq:M_of_t}. The energy $E_\sigma$ can fall anywhere in the tower of 1-meson energy eigenvalues with $k=0$. The cluster of secondary peaks are located at the remaining frequencies $\omega_{\ell g}$ for $\ell \neq \sigma$. They generally start with large amplitude at $\omega_{1g}$ and diminish in amplitude as they approach $\omega_{8g}$ (excluding $\omega_{\sigma g}$), which can similarly be explained by the Hamming distance: as $\ell$ increases from $1$ to $8$ (excluding $\sigma$), the dominant basis elements in the states $\ket{E_\ell}$ increase in Hamming distance from the $Z_2$ product state and overlap with the other basis states of $\ket{E_g}$, which themselves decrease in amplitude with increasing Hamming distance from the $Z_2$ product state, so that the corresponding matrix elements decrease.

The trends in the spectrum as we decrease $\beta < 0$ are captured in Figures \ref{fig:logMw_vs_w_beta-0.02}, \ref{fig:logMw_vs_w_beta-0.06}, \ref{fig:logMw_vs_w_beta-0.10}. The dominant peak shifts modestly to the right. The cluster of secondary peaks shift more rapidly to the right and spread, eventually passing into and through the dominant peak. The trends in the energy eigenvalue distribution are consistent (Figures \ref{fig:eigenvalues_beta0.02}, \ref{fig:eigenvalues_beta0.06}, \ref{fig:eigenvalues_beta0.10}). The ground state energy moves downwards. Though not explicitly labeled, the energy $E_\sigma$ transfers downward through the 1-meson energies. The separation between $E_g$ and $E_\sigma$ remains roughly constant, consistent with the dominant peak at $\omega_{8g}$ remaining at roughly the same location. The separation between $E_g$ and the remaining energies $E_\ell$ for $\ell \neq \sigma$ increases, consistent with the secondary peaks at $\omega_{\ell g}$ shifting to the right.
\section{Short-Range Meson Repulsion}
\label{sec:repulsion}

\begin{figure*}
\includegraphics[width=1\textwidth]{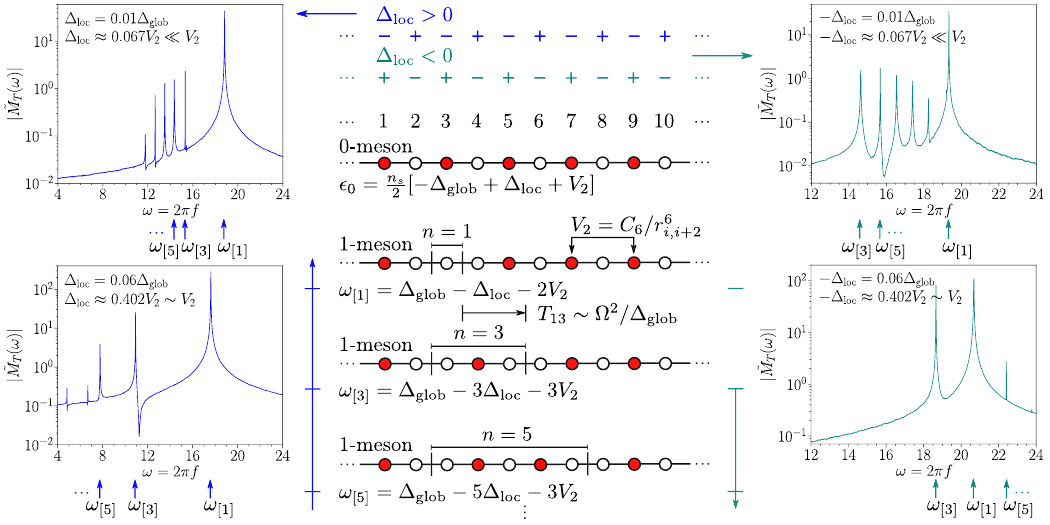}
\caption{A schematic of the short-range domain wall repulsion model. The middle section shows several atom chains, where a red circle indicates a Rydberg state and a white circle indicates a ground state. The atoms are actually configured in circles as in \Cref{fig:setup_oscillations}, but are depicted here in straight lines for convenience. The first chain shows the $Z_2$ product state $\ket{10 \hdots 10}$; it is a 0-meson state, which has no domain walls. The second, third, and fourth chains show 1-meson states, which have two $00$ domain walls, for domain sizes $n = 1, 3, 5$. The Rydberg tails are dominated by the 2nd-nearest-neighbor interaction $V_2$. We apply a staggered local detuning field $\Delta_{\mathrm{loc},j}$, defined in \Cref{eq:local_detuning}. The $Z_2$ product state is close to the metastable state for $\Delta_\mathrm{loc} > 0$; the corresponding pattern and results are shown in blue and in the left section. The $Z_2$ product state is close to the ground state for $\Delta_\mathrm{loc} < 0$; the corresponding pattern and results are shown in teal and in the right section. The vertical arrows just left and right of the middle section show the order relations for the frequencies in each case, arising from those for the energies given in \Cref{eq:fv_order_relation} and Equations~\eqref{eq:tv_order_relation_n1}-\eqref{eq:tv_order_relation_ngtr3}. The panels in the left and right sections show sample magnetization spectra for different relative values of $\Delta_\mathrm{loc}$ and $V_2$. The arrows below the panels indicate the qualitative relations between the frequencies and peaks. Our classical model of short-range meson repulsion thus shows that the dominant peak arises from the 1-meson state with $n=1$, the secondary peaks arise from the ordered 1-meson states with $n \geq 3$, the peaks shift in the direction opposite to the sign of $\Delta_\mathrm{loc}$, and the relative distribution of peaks is determined by competition between $\Delta_\mathrm{loc}$ and $V_2$.}
\label{fig:classical_energy_model}
\end{figure*}

To explain the qualitative properties of the spectral peaks, though not the precise quantitative properties (i.e. locations, spacing, etc.), we examine the classical potentials of site basis states with $0$- and $1$-meson excitations, as shown in \Cref{fig:classical_energy_model}. We consider 1-meson states with $00$ domain walls only, since $11$ domain walls are suppressed by the Rydberg blockade. Let $n$ be the domain size; it must be odd given our restriction to $00$ domain walls. The next-nearest-neighbor interaction $V_2$ characterizes the leading effect of the Rydberg tails; we ignore higher order neighbor interactions, which are further suppressed. We note that our input parameter range implies $\Delta_\mathrm{glob} \gg V_2$ and $\Delta_\mathrm{glob} \gg | \Delta_\mathrm{loc} |$. In short, the qualitative properties of the spectral peaks arise due to competition between confinement from the local detuning $\Delta_\mathrm{loc}$ and short-range domain wall repulsion from the two-site Rydberg tails $V_2$. 

We first consider the case $\beta \propto \Delta_\mathrm{loc} > 0$, shown in blue in \Cref{fig:classical_energy_model}, i.e. the upper pattern in the middle section and the outcomes in the left section. The initial state is dominated by the $Z_2$ product state $\ket{10 \hdots 10}$, shown in the first chain in \Cref{fig:classical_energy_model}, which is close to the metastable state. The $Z_2$ product state is a 0-meson state, i.e. it has no domain walls and thus domain size $n = 0$. The classical potential $\epsilon_0 \approx E_m$ of the 0-meson state is the zero point from which we will measure the 1-meson potentials, and is given by
\begin{equation}
\epsilon_0 = \frac{n_s}{2} [ -\Delta_\mathrm{glob} + \Delta_\mathrm{loc} + V_2 ] \, ,
\end{equation} 
where the first two terms arise from the detuning fields coupling to the $n_s / 2$ sites in Rydberg states and the third term arises from the $n_s / 2$ pairs of two-site Rydberg interactions. A 1-meson state with domain size $n = 1$ is shown in the second chain in \Cref{fig:classical_energy_model}. It has classical potential
\begin{equation}
\epsilon_{[1]} = \epsilon_0 + \Delta_\mathrm{glob} - \Delta_\mathrm{loc} - 2 V_2 \, ,
\end{equation}
since the state has one fewer site in the Rydberg state to couple to the detuning fields and two fewer pairs of two-site Rydberg interactions compared to the 0-meson state. Two larger 1-meson states with domain sizes $n = 3, 5$ are shown in the third and fourth chains in \Cref{fig:classical_energy_model}. A 1-meson state with $n = 3, 5, \hdots, n_s - 1$ has classical potential
\begin{equation}
\epsilon_{[n \geq 3]} = \epsilon_0 + \Delta_\mathrm{glob} - n \Delta_\mathrm{loc} - 3 V_2 \, ,
\end{equation}
since it has one fewer site in the Rydberg state to couple to the detuning fields, $(n-1)/2$ sites with reversed local detuning sign, and three fewer pairs of two-site Rydberg interactions compared to the 0-meson state. The classical potentials satisfy the order relations
\begin{equation}
\epsilon_{[1]} > \epsilon_{[3]} > \epsilon_{[5]} > \hdots > \epsilon_{[n_s - 1]} \, ,
\label{eq:fv_order_relation}
\end{equation}
and there is no order relation between $\epsilon_0$ and $\epsilon_{[n]}$. We emphasize that $\epsilon_{[1]}$ and $\epsilon_{[n \geq 3]}$ have different coefficients in the $V_2$ term, which gives $\epsilon_{[1]}$ a distinct status. 
The potential $\epsilon_{[1]}$ in the case $\beta > 0$ here is the counterpart of the energy $E_8$ in \Cref{subsec:numerical:metastable}. 
The potentials have associated frequencies $\omega_{[n]} = \epsilon_{[n]} - \epsilon_0$. The frequency order relations are shown by the blue vertical arrow to the left of the chains in \Cref{fig:classical_energy_model}. We note that $\omega_{[n]} \sim -n \Delta_\mathrm{loc}$. 

This classical analysis can explain the qualitative properties of the spectral peaks, captured in the panels on the left in \Cref{fig:classical_energy_model}: 
(1) the distinct dominant peak arises from the frequency $\omega_{[1]}$, since the corresponding state is Hamming distance $1$ away from the $Z_2$ product state; 
(2) the secondary peaks arise from the frequencies $\omega_{[n \geq 3]}$, and these frequencies decrease with increasing domain size $n$; 
(3) the secondary peaks decrease in amplitude with decreasing frequency $\omega_{[n \geq 3]}$ and thus increasing domain size $n$, since increasing the domain size produces states that are a greater Hamming distance $n$ from the $Z_2$ product state; 
(4) as $\beta > 0$ increases, the peaks shift to lower frequencies and spread, since $\omega_{[n]} \sim -n \Delta_\mathrm{loc}$; 
(5) the secondary peaks shift more strongly with $\Delta_\mathrm{loc}$ than the dominant peak, since $\omega_{[n]} \sim -n \Delta_\mathrm{loc}$; and 
(6) the spectra have at most $8$ major peaks, since the chain has $n_s = 16$ sites and $n$ only takes odd values due to the Rydberg blockade.

The relative distribution of the dominant and secondary peaks remain the same for all values of $\Delta_\mathrm{loc}$ and $V_2$, as observed in the left panels of \Cref{fig:classical_energy_model} and understood from the frequency order relations. In other words, the secondary peaks lie to the left of the dominant peak, and shift and spread to the left as $\Delta_\mathrm{loc} > 0$ increases. 

We next consider the case $\beta \propto \Delta_\mathrm{loc} < 0$, shown in teal in \Cref{fig:classical_energy_model}, i.e. the lower pattern in the middle section and the outcomes in the right section. The initial state is again dominated by the $Z_2$ product state, which is now close to the ground state. The $Z_2$ product state has classical potential $\epsilon_0 \approx E_g$. The expressions that result for the classical potentials and frequencies are the same as in the metastable case. However, since $\Delta_\mathrm{loc} < 0$ now, they satisfy different order relations. The classical potentials satisfy the order relations
\begin{align}
\epsilon_0 &< \epsilon_{[n]} \, , \label{eq:tv_order_relation_n1} \\
\epsilon_{[3]} &< \epsilon_{[5]} < \hdots < \epsilon_{[n_s - 1]} \, , \label{eq:tv_order_relation_ngtr3}
\end{align}
and there is no order relation between $\epsilon_{[1]}$ and $\epsilon_{[n \geq 3]}$. 
The potential $\epsilon_{[1]}$ in the case $\beta < 0$ here is the counterpart of the energy $E_\sigma$ in \Cref{subsec:numerical:ground}. 
The frequency $\omega_{[1]}$ can thus fall anywhere within the set of frequencies $\omega_{[n \geq 3]}$. The frequency order relations are shown by the teal vertical arrow to the right of the chains in \Cref{fig:classical_energy_model}. 

The classical analysis applies similarly to the ground case, producing some different results, captured in the panels on the right in \Cref{fig:classical_energy_model}: 
($2'$) the secondary peaks arise from the frequencies $\omega_{[n \geq 3]}$, and these frequencies increase with increasing domain size $n$; 
($3'$) the secondary peaks decrease in amplitude with increasing frequency $\omega_{[n]}$ and thus increasing domain size $n$, since increasing the domain size produces states that are a greater Hamming distance $n$ from the $Z_2$ product state; and 
($4'$) as $\beta < 0$ decreases, the peaks shift to higher frequencies and spread, since $\omega_{[n]} \sim -n \Delta_\mathrm{loc}$.

The relative distribution of the dominant and secondary peaks depend on the relative strengths of $\Delta_\mathrm{loc}$ and $V_2$, as observed in the right panels of \Cref{fig:classical_energy_model} and understood from the frequency order relations. The relative values fall in three broad categories. Small local detuning $(-\Delta_\mathrm{loc} \ll V_2)$: the secondary peaks lie in a cluster to the left of the dominant peak and are closely spaced, as shown in the top right panel in \Cref{fig:classical_energy_model} and previously in \Cref{fig:logMw_vs_w_beta-0.01}. Intermediate local detuning $(-\Delta_\mathrm{loc} \sim V_2)$: the secondary peaks have shifted and spread to the right, and now lie across the dominant peak with wider spacing, some of the peaks passing through the dominant peak and some having merged with it, as shown in the bottom right panel in \Cref{fig:classical_energy_model} and previously in Figures \ref{fig:logMw_vs_w_beta-0.02} and \ref{fig:logMw_vs_w_beta-0.06}. Large local detuning $(-\Delta_\mathrm{loc} \gg V_2)$: the secondary peaks have shifted and spread further to the right, and now lie to the right of the dominant peak with wider spacing, some having merged with it, as shown previously in \Cref{fig:logMw_vs_w_beta-0.10}.

We can relate the classical potentials for the 1-meson states to the neutral atom version of the potential $U_0(n)$ in the 1-meson model in \Cref{eq:1meson}. In particular, the potential is simply $U_0 (n) = \epsilon_{[n]}$. This potential has several distinctive features. First, it only involves odd domain sizes $n = 1, 3, \hdots, n_s - 1$. Second, it has a term $-\Delta_\mathrm{loc} n$ for all $n$, but different terms $-2 V_2$ for $n = 1$  and $-3 V_2$ for $n \geq 3$. The peak arising from $n = 1$ is thus qualitatively different from the peaks arising from $n \geq 3$. In contrast, in the nearest-neighbor Ising spin chain, the potential supports all domain sizes $n$ and has the form $U(n) \propto h_z n$ for all $n$ (\Cref{sec:quasiparticle}). 

The classical model is thus accurate when the kinetic energy term $T_{nn'}$ is much less than the potential different $\Delta U_0$ in the 1-meson model in \Cref{eq:1meson}. In that case, the 1-meson model becomes $U_0 (n) \phi_\ell (n) \approx (\varepsilon_\ell / 2) \phi_\ell (n)$. The largest kinetic energy term has scaling $T_{n,n+2} \sim \Omega^2 / \Delta_\mathrm{glob}$, ignoring smaller effects from $V_1$ and $V_2$, since expanding a domain wall from size $n$ to $n+2$ involves two Rabi transitions, a second order process, and the intermediate state with 00 sites yields a global detuning penalty, as shown by the term $T_{13}$ in \Cref{fig:classical_energy_model}. The potential difference from size $n$ to $n+2$ has scaling $\Delta U_0 \sim  \Delta_\mathrm{loc}$, again ignoring smaller effects from $V_1$ and $V_2$. The ratio of the potential to kinetic terms is $\sim \Delta_\mathrm{loc} \Delta_\mathrm{glob} / \Omega^2 = \beta \alpha^2$, and thus our classical model is an accurate approximation for $\beta \alpha^2 \gg 1$. The competition is manifested visually in the energy eigenvalues of $\tilde{H}_0$ (approximately those of $H$): $T_{n,n+2}$ scales with the energy range, or curvature, of the 1-meson bands across momentum $k$, and $U_0 (n)$ scales with the energy difference $\omega_{mg} \propto \Delta_\mathrm{loc}$ between the metastable and ground states. For instance, the case $\beta = \pm 0.1$ gives $\beta \alpha^2 = 1.6$ and is shown in \Cref{fig:eigenvalues_beta0.10}, whereas the case $\beta = \pm 0.01$ gives $\beta \alpha^2 = 0.16$ and is shown in \Cref{fig:eigenvalues_beta0.01}; the case $\beta = \pm 0.1$ is closer to the classical limit, and thus the 1-meson bands are more flat and the ground and metastable states have larger separation. However, the classical energy model is also approximately valid for larger kinetic energy, since the leading effect of the latter is insensitive to the domain size, i.e. it shifts the 1-meson energies by a term largely independent of the domain size~\cite{Rutkevich:1999}.

We can compare our work to that of Lagnese et al.~\cite{Lagnese:2023}, who performed a supplementary study of metastable oscillations in neutral atom chains with Rydberg excitations. The features we observe for the metastable case in \Cref{fig:logMw_vs_w_beta0.10} for $\beta = 0.10$ appear similar to those observed in Ref.~\cite{Lagnese:2023} Figure S1b for $\beta = 0.125$. Furthermore, the features we observe for the ground case in \Cref{fig:logMw_vs_w_beta-0.10} for $\beta = -0.10$ appear similar to those observed in Ref.~\cite{Lagnese:2023} Figure S1a for $\beta = -0.125$. Lagnese et al.~\cite{Lagnese:2023} interpreted their observations in analogy with their main results for the ferromagnetic Ising spin chain. However, in our detailed investigations with neutral atom chains that include long-range Rydberg interactions, we found that the peaks have different structure, exhibit different shifting behavior, and arise from different microphysics, all distinct to these antiferromagnetic Rydberg atom systems.
\section{Experimental Accessibility}
\label{sec:experiments}

In this section, we investigate the experimental accessibility of the oscillation spectra. The hardware constraints on a prototypical system include~\cite{aquila2023quera}: a maximum simulation time limit $t = 4$ $\mu$s; a minimum atomic spacing $a_\mathrm{min} = 4.0$ $\mu$m; a field of view $(F_x, F_y) = (75.0, 76.0)$ $\mu$m; a Rabi frequency range $\Omega \in [0.00, 15.8]$ rad/$\mu$s; and a global detuning range $\Delta_\mathrm{glob} \in [-125.0, 125.0]$ rad/$\mu$s. 

In \Cref{sec:numerical}, we presented a numerical setup to study metastable long-lived oscillations, with the following input parameters: $\Omega / 2\pi = 1.0$ MHz ($R_b = 9.76$ $\mu$m); $a = 5.42$ $\mu$m; $\Delta_\mathrm{glob} / 2\pi = 4.0$ MHz; and $\Delta_\mathrm{loc} / 2\pi \in [0.0008, 0.06]$ MHz. 

An experimental protocol would consist of a specified atom configuration and a waveform sequence adapted to it, consisting of a preparation stage and a quench stage. The preparation stage would involve adiabatic construction of the ground state of $H$ for $\beta_\mathrm{prep} = -0.001$. The quench stage would produce $H$ with the waveforms in our numerical setup. The main chain of $n_s = 16$ atoms would be arranged along a closed path to achieve periodic boundary conditions. If arranged in a square-like path, the height and width would be $d \sim 4a \approx 22$ $\mu$m, well below $F_x$ and $F_y$. The setup could also use ancilla atoms to aid the preparation stage, which the field of view can readily accommodate. 

We estimate a preparation duration $t_\mathrm{prep} \approx 2.0$ $\mu$s based on previous experiments (e.g.~\cite{Bernien:2017,Bluvstein:2021}), and thus obtain a maximum post-quench duration $t_\mathrm{evolve} \approx 2.0$ $\mu$s to run the protocol within the simulation time limit. The post-quench duration presents the main obstacle to obtaining identifiable features in the spectra. However, the relevant timescale to resolve the spectra is not $t_\mathrm{evolve}$, but the (dimensionless) effective counterpart $\Omega' t_\mathrm{evolve} / 2\pi$ that is scaled by the Rabi frequency $\Omega'$, since Rabi oscillations set the dynamic range. A larger Rabi frequency $\Omega'$ will increase the effective duration. It will also decrease the blockade radius to $R'_b = R_b / (\Omega' / \Omega)^{1/6}$, and thus require an increased atom separation $a' = (\Omega' / \Omega)^{1/6} a$ to achieve the same $R_b / a$. The spatial extent of the configuration with this larger separation must still lie within the field of view.

A Rabi frequency $\Omega' / 2\pi = 2.0$ MHz would yield $\Omega' t_\mathrm{evolve} / 2\pi = 4.0$. The resulting Fourier spectra are shown by the lowest opacity curves for $t_\mathrm{Stop} = 4.0$ given $\Omega / 2\pi = 1.0$ MHz in Figures \ref{fig:oscillation_properties} and \ref{fig:spectra_beta_large}. The spectra only partially resolve the dominant peak and do not resolve the cluster of secondary peaks. At most, one can observe the shifting and rolling of the broadened dominant peak with changing local detuning parameter. 

In contrast, a Rabi frequency $\Omega' / 2\pi = 20.0$ MHz would yield $\Omega' t_\mathrm{evolve} / 2\pi = 40$. For comparison, experiments investigating topological spin liquids performed a quench to $\Omega' / 2\pi = 20.0$ MHz for $\Delta t \lesssim 40$ ns~\cite{Semeghini:2021}. This Rabi frequency would require an atom separation $a' = 8.93$ $\mu$m to achieve the same $R_b / a$, and thus height and width $d \sim 4a \approx 36$ $\mu$m, still well below $F_x$ and $F_y$. The resulting Fourier spectra are shown by the intermediate opacity curves for $t_\mathrm{Stop} = 40.0$ given $\Omega / 2\pi = 1.0$ MHz in Figure \ref{fig:oscillation_properties}. The spectra sufficiently resolve the cluster of secondary peaks. 

The experiments also provide an opportunity to study the oscillation spectra as a function of system size. If the main chain is arranged in a rectangular path, the fields of view could accommodate upwards of $n_x \sim \lfloor F_x / a \rfloor = 13$ atoms and $n_y \sim \lfloor F_y / a \rfloor = 14$ atoms, or a chain with $n_s \sim 2 n_x + 2 n_y = 54$ atoms. A path that is more meandering could accommodate an even larger number of atoms. If the setup also used ancilla atoms to aid the preparation stage, it could reduce the number of possible atoms in the main chain depending on the placement. The oscillation timescales are largely independent of the system size since they are mainly set by the 1-meson energy gap, which is independent of it as well. In contrast, the preparation stage becomes increasingly challenging with increasing system size; however, the use of a larger Rabi frequency can help offset this in equal measure. In addition, the local detuning can be used to circumvent the second order gap crossing, and thus increase the state preparation fidelity.
\section{Conclusion and Outlook}
\label{sec:conclusion}

In this paper, we studied long-lived oscillations of metastable and ground states using neutral atoms with long-range Rydberg interactions. Our work examined setups accessible to near-term neutral atom experiments. We computed the Fourier spectra of quasiparticle oscillations initiated by metastable and ground states, and identified spectral features to distinguish between these two cases. In particular, we discovered features distinct to our antiferromagnetic system compared to ferromagnetic Ising chains, and explained them in terms of confinement and short-range domain wall repulsion effects. 

Our results show that the features of metastable and ground state oscillation spectra contain fingerprints of the underlying confining potential. As a consequence of this dependence on the microscopic details, it is crucial that one examines the underlying microscopic model through simulations before drawing universal conclusions about metastability. This underscores the importance of nascent quantum simulators in addressing metastability phenomenology.

Our work motivates several further studies. Experiments on 1D neutral atom systems, including analog quantum simulators~\cite{Henriet:2020,aquila2023quera}, can probe the oscillation spectra that we observed, motivated by our parameter range (\Cref{sec:numerical}) and accessibility discussion (\Cref{sec:experiments}). However, as noted in \Cref{sec:experiments}, a detailed investigation of the secondary peaks will require intense Rabi pulses to enhance the effective simulation duration or improvements in the coherence time to extend the physical simulation duration. Long-lived oscillations can also be studied on 2D lattices~\cite{Scholl:2021,Ebadi:2021}; this could access novel metastability phenomenology that only emerges in 2D~\cite{Voloshin:2004}. In 2D systems that grow beyond $\mathcal{O}(100)$ sites, classical simulation becomes increasingly intractable as a tool to study metastable oscillations~\cite{Daley:2023,Shaw:2024}. Early quantum devices could thus be an exclusive window into the physics of higher-dimensional quantum field theories.
\section*{Computing Resources}

We ran our simulations on the high-performance computing system Perlmutter at the National Energy Research Scientific Computing Center (NERSC) based at Lawrence Berkeley National Laboratory~\cite{perlmutter2023nersc} and used the Bloqade software package developed by QuEra Computing~\cite{bloqade2023quera}.

\begin{acknowledgments}
This research was supported by the U.S. Department of Energy (DOE) under Contract No. DE-AC02-05CH11231, through the National Energy Research Scientific Computing Center (NERSC), an Office of Science User Facility located at Lawrence Berkeley National Laboratory. R.V.B. was supported by the Office of Science, Office of Advanced Scientific Computing Research (ASCR) Exploratory Research for Extreme-Scale Science. The numerical study was performed on the high-performance computing system Perlmutter, a NERSC resource, using NERSC award DDR-ERCAP0030190.
\end{acknowledgments}


\bibliography{references} 

\end{document}